\documentclass[aps,prd,twocolumn,showpacs,preprintnumbers,nofootinbib,amsmath,amssymb]{revtex4}

\usepackage{graphicx}
\usepackage{dcolumn}
\usepackage{bm}
\usepackage{amsmath}
\usepackage{braket}
\usepackage[normalem]{ulem}
\usepackage[hidelinks]{hyperref}
\usepackage{epstopdf}
\usepackage{placeins}
\usepackage[normalem]{ulem}
\usepackage{multirow}
\usepackage{makecell}

\usepackage{tikz}
\usetikzlibrary{arrows.meta}
\usetikzlibrary{positioning}
\usetikzlibrary{decorations}
\usetikzlibrary{decorations.markings}
\tikzset{
	lattice point/.style={
		draw,
		fill,
		circle,
		minimum size=1.0mm,
		inner sep=0,
	},
	intermediate point/.style={
		draw,
		fill,
		circle,
		minimum size=0.0mm,
		white,
		inner sep=0,
	},    
}

\newcommand{\beq}{\begin{eqnarray}}
\newcommand{\eeq}{\end{eqnarray}}
\newcommand{\beqnn}{\begin{eqnarray*}}
\newcommand{\eeqnn}{\end{eqnarray*}}

\def\spose#1{\hbox to 0pt{#1\hss}}
\def\ltapprox{\mathrel{\spose{\lower 3pt\hbox{$\mathchar"218$}}
 \raise 2.0pt\hbox{$\mathchar"13C$}}}

\begin{document}

\title{Large-$N$ expansion and $\theta$-dependence of $2d$ $CP^{N-1}$ models
beyond the leading order}

\author{Mario Berni}
\email{mario.berni@pi.infn.it}
\affiliation{Universit\`a di Pisa and INFN Sezione di Pisa,\\ 
Largo Pontecorvo 3, I-56127 Pisa, Italy
}

\author{Claudio Bonanno}
\email{claudio.bonanno@pi.infn.it}
\affiliation{Universit\`a di Pisa and INFN Sezione di Pisa,\\ 
Largo Pontecorvo 3, I-56127 Pisa, Italy
}

\author{Massimo D'Elia}
\email{massimo.delia@unipi.it}
\affiliation{Universit\`a di Pisa and INFN Sezione di Pisa,\\ 
Largo Pontecorvo 3, I-56127 Pisa, Italy
}

\date{\today}

\begin{abstract}
We investigate the $\theta$-dependence 
of 2-dimensional $CP^{N-1}$ models in the large-$N$ limit by lattice simulations.
Thanks to a recent algorithm proposed by M.~Hasenbusch to improve the 
critical slowing down of topological modes, 
combined with simulations at imaginary values of $\theta$,
we manage to determine the vacuum energy
density up the sixth order in $\theta$ and
up to $N = 51$.
Our results support analytic predictions,
which are known up to the next-to-leading term in $1/N$
for the quadratic term in $\theta$ (topological susceptibility), 
and up to the leading term for the quartic coefficient $b_2$. 
Moreover, we give a numerical estimate of further 
terms in the $1/N$ expansion for both quantities, pointing out that
the $1/N$ convergence for the $\theta$-dependence 
of this class of models is particularly slow.
\end{abstract}

\maketitle

\section{Introduction}
\label{section_intro}

The existence of field configurations with non-trivial topology
characterize the non-perturbative properties
of QCD and QCD-like models, leading in particular
to a non-trivial dependence on a 
possible coupling
to the topological charge operator $q(x)$,
the so-called $\theta$ parameter.
A non-zero $\theta$ implies an additional
factor $\exp(i \theta Q)$ in the path-integral of the theory,
where $Q = \int d^4x \, q(x)$ is the global topological charge
(winding number); 
since $Q$ is integer-valued
for field configurations decaying fast enough at infinity, 
the theory is invariant under $\theta \to \theta + 2 \pi$,
so that $\theta$ behaves as an angular variable.
For small values of 
$\theta$, the free energy (vacuum energy)
density $f(\theta)$ can be Taylor expanded
around $\theta = 0$, a common parameterization 
being~\cite{vicari_review_QCD_CPN}
\begin{equation}
f(\theta) - f(0) = \frac{1}{2}\chi \theta^2\left(1+\sum_{n=0}^{\infty}b_{2n}\theta^{2n}\right) \, .
\label{taylexp}
\end{equation}
The expansion contains only even terms because $\theta$ breaks 
CP-symmetry explicitly and the theory is CP-invariant at $\theta = 0$.
The quadratic coefficient
$\chi$ is the topological susceptibility and is related to the second cumulant of the topological charge distribution at $\theta=0$, while the coefficients
$b_{2n}$ are related to higher-order cumulants of this distribution.

Since $\theta$-dependence is connected
to intrinsically non-perturbative properties of Quantum Field Theories,
a numerical approach based on lattice Monte Carlo
simulations is the natural first-principle approach to 
its investigation. However, various analytical strategies
permit to obtain useful information, at least 
in certain limits.

Semiclassical approaches to $\theta$-dependence consider 
classical configurations with non-trivial topology,
like instantons and anti-instantons, and compute the path-integral
by integrating fluctuations around such class of configurations.
One usually considers configurations with just one instanton
or anti-instanton, so that the whole information is contained
in the single-instanton effective action. In this way, one obtains
results valid for an ensemble of independent (non-interacting)
instantons and anti-instantons (dilute instanton gas approximation
or DIGA), leading to a universal dependence of the free energy 
$\theta$:
\beq\label{DIGA}
f(\theta)  - f(0) = \chi (1 - \cos \theta) \, .
\eeq
This approximation is expected to work well at least in some regimes,
like for $SU(N)$ Yang-Mills theories in the deconfined, high-temperature
phase, where the typical instanton effective action gets large (because
of asymptotic freedom) and topological charge fluctuations become 
rare and dilute.

On the other hand, 
DIGA is known to break down when instanton interactions cannot be neglected,
like in the confined phase of QCD and similar models. 
In this regime, an alternative approach is represented by an expansion in
the inverse of the number of field components, 
i.e.~in $1/N$ for $SU(N)$ Yang-Mills theories or for 
$CP^{N-1}$ models.
Under very general assumptions, like requiring the existence 
of a non-trivial
dependence on $\theta$, large-$N$ expansion leads 
at least to semi-quantitative
insights, like the prediction that the coefficients
$b_{2n}$ in Eq.~(\ref{taylexp}) be suppressed as 
$1/N^{2n}$~\cite{witten0, witten2}, 
which for $SU(N)$ Yang-Mills theories has been checked by lattice simulations during
the last few years~\cite{DelDebbio:2002xa,DElia:2003zne,Giusti:2007tu,first_article_theta_immaginario_SUN,Ce:2015qha,theta_dep_SU3_d_elia,Bonati:2013tt,SU(N)_large_N_limit,Bonati:2018rfg}.

The case of $CP^{N-1}$ models in two dimensions, which is the subject 
of the present investigation, is special, because the large-$N$
expansion permits to obtain for them, as for other vector-like
models, also quantitative predictions.
Leading order computations are available for $\chi$ \cite{cpn_article_confinement}
and for all $b_{2n}$ coefficients \cite{SU(N)_large_N_limit,cpn_b_2_theo}, and even next-to-leading 
corrections are known for the topological susceptibility~\cite{calcolo_e_2}.
Despite the fact that numerical simulations for 
these models are less demanding than those for $SU(N)$ 
Yang-Mills theories, lattice results have failed
up to now to provide a clear confirmation of 
these quantitative analytical predictions, apart
from the case of the leading $1/N$ term for $\chi$ \cite{susc_top_cpn,vicari_simul_temp_cpn,critical_slowing_down_review, fighting_slowing_down_cpn}.
Numerical results appeared in some cases to be even inconsistent
with next-to-leading predictions for $\chi$~\cite{critical_slowing_down_review, fighting_slowing_down_cpn}.
A recent investigation~\cite{Bonanno:2018xtd} pointed out that at least 
the consistency can be recovered once one takes into 
account further terms in the $1/N$ expansion.
The purpose of the present investigation is to go beyond
the simple consistency, trying to achieve a quantitative 
agreement between lattice computations and analytical predictions,
at least for the next-to-leading correction to $\chi$
and for the leading term in the $b_2$ coefficient: 
in Ref.~\cite{Bonanno:2018xtd} it was suggested that, in order to do so,
one should explore $N = 50$ or larger.
At the same time, we would like to achieve a numerical estimate
of the further terms in the $1/N$ expansion for both quantities.

In order to achieve our goal, we have pushed our investigation
up to $N = 51$,  where however standard
algorithms face severe critical slowing down problems
in the decorrelation of the topological 
charge~\cite{critical_slowing_down_review}, 
which can only partially be ameliorated by numerical strategies
like simulated (or parallel) tempering in the coupling of 
the theory~\cite{vicari_simul_temp_cpn,Bonanno:2018xtd}.
For this reason, we have decided to adopt an algorithm
recently introduced by M.~Hasenbusch in 
Ref.~\cite{fighting_slowing_down_cpn}, in which 
simulations with open and periodic boundary conditions
are smartly combined in a parallel tempering framework.
In addition to that, following the same strategy
adopted in Ref.~\cite{Bonanno:2018xtd},
we will assume analyticity 
around $\theta = 0$ and exploit simulations performed 
at imaginary values of $\theta$ in order to improve the 
signal-to-noise ratio, something which turns out to be essential
in order to achieve a precise 
determination of the 
higher-order cumulants of $Q$.

The paper is organized as follows. 
In Section~\ref{cpn_continuum} we 
provide a concise review of 
$CP^{N-1}$ models and of large-$N$
predictions for their $\theta$-dependence.
In Section~\ref{section_setup} we give details
about the lattice discretization and the numerical 
algorithm employed in our study.  
Numerical results and their analysis within the framework of 
the $1/N$ expansion are presented in Section~\ref{section_results}.
Finally, in Section~\ref{section_conclusions}, we give our conclusions.

\section{$CP^{N-1}$ models and their $\theta$-dependence in the large-$N$ limit}
\label{cpn_continuum}

$CP^{N-1}$ models in two space-time dimensions
share many properties with Yang-Mills theories: apart
from $\theta$-dependence, they are also 
characterized by confinement of fundamental matter fields; for this 
reason they have represented a theoretical test-bed for the 
study of non-perturbative physics in gauge theories since long~\cite{witten,
cpn_article_confinement,cpn_general_properties,advanced_topics_QFT}.

The elementary fields belong to the projective space of 
$N$-component complex vectors. The projective conditions
is enforced by normalizing the modulus of the vectors fields
to one and by writing an action which is independent of 
the residual arbitrary local phase factor of the fields. 
In some formulations, such as the one considered in our study,
this is rephrased by introducing an auxiliary and non-propagating
abelian gauge field $A_\mu$, so that the arbitrary local phase is gauged
away with the advantage of having an action 
quadratic in the fields. In particular the Euclidean action, already including 
the $\theta$-term, reads
\beq
S(\theta)= \int \left[\frac{N}{g} \bar{D}_\mu \bar{z}(x) D_\mu z(x) -i \theta q(x) \right]d^2x \, ,
\eeq
where $z$ is a complex $N$-component scalar field satisfying
$\bar{z}(x)z(x)=1$, $D_\mu$ is the usual $U(1)$ covariant derivative, $g$ is
the 't Hooft coupling, which is kept fixed as $N \to \infty$, and
\beq
Q=\int q(x) d^2x =\frac{1}{4\pi}\epsilon_{\mu\nu}\int F_{\mu\nu}(x) d^2x
\eeq
is the global topological charge. The free energy (or vacuum energy) density
is defined, using the path-integral formulation of the theory, as
\beq
f(\theta) \equiv -\frac{\log Z(\theta)}{V} = -\frac{1}{V}\log \int [dA][d\bar{z}][dz] e^{-S(\theta)} \,,
\eeq
where $V$ is the 2$d$ space-time volume. From this expression it is clear
that the parameters entering the Taylor expansion of $f(\theta)$ 
around $\theta = 0 $, see Eq.~(\ref{taylexp}), can be related
to the cumulants $k_n$ of the path-integral distribution of the 
topological charge, $P(Q)$, computed at $\theta = 0$:
\beq
\chi &=& \frac{1}{V} k_2\bigg\vert_{\theta=0} =
\frac{1}{V}\braket{Q^2}\bigg\vert_{\theta=0}, \nonumber \\
b_2 &=& -\frac{k_4}{12 \text{ } k_2}\bigg\vert_{\theta=0} =
\frac{-\braket{Q^4}+3\braket{Q^2}^2}{12 \braket{Q^2}}\bigg\vert_{\theta=0}\, , \label{definition_chi_b_2n} \\
b_4 &=& \frac{k_6}{360 \text{ } k_2}\bigg\vert_{\theta=0} =
\frac{\braket{Q^6}-15\braket{Q^4}\braket{Q^2}+30\braket{Q^2}^3}{360 \braket{Q^2}}\bigg\vert_{\theta=0}\, . \nonumber
\eeq
Large-$N$ arguments predict\footnote{Notice that such predictions are valid for the vacuum, while at finite temperature the $\theta$-dependence could be different, see, e.g., Refs.~\cite{Bolognesi:2019rwq, Flachi:2019jus, Fujimori:2019skd} for a recent discussion.}
$\chi=\bar{\chi}N^{-1}+O(N^{-2})$ and
$b_{2n}=\bar{b}_{2n}N^{-2n}+O(N^{-2n-1})$. On a more quantitative level, 
one can show that
\cite{calcolo_e_2, SU(N)_theta_dep_manca, SU(N)_large_N_limit, cpn_large_N}
\beq \label{large_N}
\xi^2 \chi &=& \frac{1}{2\pi N}+\frac{e_2}{N^2}+O\left(\frac{1}{N^3}\right) \, ,\\
b_2 &=& -\frac{27}{5}\frac{1}{N^2}+O\left(\frac{1}{N^3}\right) \, , \label{b2largen}\\
b_4 &=& -\frac{25338}{175}\frac{1}{N^4}+O\left(\frac{1}{N^5}\right) \, ,
\eeq
where the length scale $\xi$ appearing in
Eq.~(\ref{large_N}) is the second moment correlation length, defined as:
\beq \label{def_xi}
\xi^2 \equiv \frac{1}{\int G(x)d^2x}\int G(x) \frac{\vert x\vert^2}{4} d^2 x \,,
\eeq
with
\beq\label{projector_definition}
G(x) \equiv \braket{P_{ij}(x)P_{ij}(0)}-\frac{1}{N}, \ P_{ij}(x) \equiv z_i(x) \bar{z}_j(x) \,.
\eeq
The next-to-leading correction to $\xi^2 \chi$
is the result of a non-trivial analytic computation performed in
Ref.~\cite{calcolo_e_2}, leading to the prediction $e_2 \simeq -0.0605$.
Numerical simulations have fully confirmed the leading order behavior 
of 
$\xi^2 \chi$~\cite{MC_simulation_cpn, fighting_slowing_down_cpn, susc_top_cpn,
renormalization_lattice_top_charge_1}, however so far they 
have been elusive in confirming the prediction for $e_2$:
many numerical works on the
$CP^{N-1}$ theories show a deviation from the leading term which appears to be
of opposite (positive) sign, and only recently the hypothesis has been
made that this could be due to a poor convergence of the series
due to quite large next-to-next-to-leading-order (NNLO) 
contributions~\cite{Bonanno:2018xtd}.  
Also for the
$O(\theta^4)$ coefficient $b_2$, consistency with the prediction
in Eq.~(\ref{b2largen}) is found only assuming large
NNLO corrections~\cite{Bonanno:2018xtd}.

\section{Numerical Setup}
\label{section_setup}

In the following we describe various aspects of the numerical methods
used in this investigation, starting from the discretization adopted
for the path-integral and for the topological observables, then
describing the strategy based on the introduction of an imaginary
$\theta$ term and on analytic continuation, and finally discussing 
the application of the Hasenbusch algorithm~\cite{fighting_slowing_down_cpn}
to our numerical setup.

\subsection{Lattice discretization}\label{subsection_lattice_action}

The theory has been put on a square lattice of size
$L$ and, even if the updating algorithm considers different
kinds of boundary conditions at the same time in a parallel
tempering framework, average values of observables have been
computed only in the case of  periodic boundary conditions (p.b.c.).
We have adopted 
the tree-level Symanzik-improved lattice discretization
for the non-topological part of the action~\cite{MC_simulation_cpn}
\begin{multline}\label{Symanzik_improved_lattice_action_cpn}
S_L = - 2N\beta_L\sum_{x,\mu} \left \{ c_1 \Re\left[\bar{U}_\mu(x)\bar{z}(x+\hat{\mu})z(x)\right] \right. \\
\left. + c_2 \Re\left[\bar{U}_\mu(x+\hat{\mu})\bar{U}_\mu(x)\bar{z}(x+2\hat{\mu})z(x)\right] \right\} \, ,
\end{multline}
where $c_1=4/3$, $c_2=-1/12$, $\beta_L \equiv 1/g_L$ is the inverse 
bare coupling and
$U_\mu(x)$ are the $U(1)$ elementary parallel transporters.
The coefficients $c_1$ and $c_2$ are chosen so as to cancel logarithmic corrections to the leading 
$O(a^2)$ approach to the continuum limit, where $a$ is the 
lattice spacing.

The continuum limit is achieved, by asymptotic freedom, 
as $\beta_L \to \infty$. In this limit the lattice correlation
length $\xi_L$ diverges as $1/a$;
$\xi_L$ is defined as usual in terms of two-point
correlation functions~\cite{original_article_corr_length}:
\beq \label{def_xi_L}
\xi_L^2 = \frac{1}{4\sin^2\left(k/2\right)}\left[ \frac{\tilde{G}_L(0)}{\tilde{G}_L(k)}-1 \right] \, ,
\eeq
where $\tilde{G}_L(p)$ is the Fourier transform of $G_L$, which is 
the discretized version of the two-point correlator of $P$ defined 
in Eq.~(\ref{projector_definition}), and $k=2\pi/L$.  
Corrections to continuum scaling can be expressed as inverse
powers of $1/\xi_L$ so that, for the adopted discretization,
the expectation value of a generic observable will scale
towards the continuum as:
\beq \label{continuum_limit_scaling_xi_L}
\braket{\mathcal{O}}_L\left(\xi_L\right) = \braket{\mathcal{O}}_{\text{\textit{cont}}} + c \, \xi_L^{-2} + 
O\left(\xi_L^{-4}\right) \,.
\eeq

Regarding the topological charge $Q$, several lattice discretizations
exist, all agreeing in the continuum limit, where a well defined
classification of relevant configurations in homotopy sectors
is recovered.
In general, any lattice discretization $Q_L$ of the topological charge
operator will be related to the continuum one by a finite
multiplicative renormalization \cite{renormalization_lattice_top_charge_1}:
\beq
Q_L = Z\left(\beta_L\right) Q \, .
\eeq
The above relation holds when one considers correlation functions
of $Q_L$,~i.e.~it is not valid configuration by configuration, where
one should instead write $Q_L = Z Q + \eta$,
where $\eta$ is noise contribution related to field fluctuations 
at the ultraviolet (UV) scale, which is stochastically independent
of the global topological background $Q$ but can lead to further
additive renormalizations as correlations with higher powers of $Q_L$ 
are considered.

Various smoothing algorithms have been commonly adopted in the
literature to dampen UV fluctuations responsible for
such renormalizations, like cooling~\cite{Berg:1981nw}, the
gradient flow~\cite{Luscher:2010iy}, or smearing; 
these procedures have been shown to be practically 
equivalent, once they are appropriately
matched to each other~\cite{Alles:2000sc, cooling_vs_gradient_flow,
Alexandrou:2015yba}. In this study we adopt cooling, because of its relative
simplicity, which consists in the sequential application of 
local modifications of the lattice fields in which the action is 
minimized locally at each step. For the purpose of smoothing, 
the minimized action can be different from the one used
to define the path-integral, our choice has been
to set $c_1=1$ and $c_2=0$ in
Eq.~(\ref{Symanzik_improved_lattice_action_cpn}) for cooling.

The most
straightforward discretization of $Q$ makes use of the plaquette operator
$\Pi_{\mu\nu}(x)$:
\beq \label{def_non_geo_charge}
Q_L=\frac{1}{4\pi}\sum_{x,\mu,\nu} \epsilon_{\mu\nu}\Im\left[\Pi_{\mu\nu}(x)\right] =\frac{1}{2\pi}\sum_x \Im\left[\Pi_{12}(x)\right] \, ,
\eeq
where, as usual,
\beq
\Pi_{\mu \nu} (x) \equiv U_\mu(x) U_\nu(x+\hat{\mu})\bar{U}_\mu(x+\hat{\nu})\bar{U}_\nu(x) \, .
\eeq
This choice leads to an analytic function of the gauge fields
which is non-integer valued and has $Z < 1$. There are alternative
definitions, known as geometric, which are always integer valued
for p.b.c. (hence they have $Z = 1$); 
one possibility~\cite{MC_simulation_cpn} is based on the link variables 
$U_\mu(x)$
\beq
Q_{U} = \frac{1}{2\pi}\sum_{x} \Im \left\{ \log \left[\Pi_{12}(x)\right] \right\} \, ,
\eeq
the other~\cite{original_article_lattice_top_charge_triangle} 
on the projector $P$ defined in Eq.~(\ref{projector_definition}). 
The geometric charge
$Q_U$ can be easily interpreted as the sum of the magnetic fluxes
(modulo $2 \pi$) going out of each plaquette, then
normalized by $2 \pi$, which is integer valued for any $2d$ compact 
manifold.

We have adopted the unsmoothed non-geometric definition $Q_L$ 
to introduce a $\theta$-term in the action, even
if this implies the presence of renormalization effects
which will be discussed below: 
$Q_L$ is linear in the fields, hence it allows to 
make use of standard efficient algorithms like over-heatbath.
As for measurements, it has been 
checked~\cite{Bonanno:2018xtd} that 
all definitions yield practically indistinguishable results after
the application of a modest amount of cooling, in particular
$O(10)$ sweeps of local minimization of each site/link variable
over the whole lattice; in any case we adopted the geometric
one (measured after 25 cooling step) which 
always yields exactly integer values.

\subsection{Imaginary-$\theta$ method}\label{subsec_imm_theta}

The coefficients $b_{2n}$ appearing in the Taylor expansion of $f(\theta)$ 
around $\theta = 0$ are observables plagued
by a large noise-to-signal ratio, especially when one tries
to determine them in terms of the topological charge distribution
at $\theta = 0$, as in Eq.~(\ref{definition_chi_b_2n}), 
since one has to measure tiny non-gaussianities 
in an almost-Gaussian distribution. A better strategy 
is to add a source term to the action,
i.e.~to consider the theory at $\theta \neq 0$,
and study the dependence of lower cumulants on $\theta$,
which contains the relevant information on the higher order cumulants.
This is not possible in practice, because the theory at non-zero
$\theta$ has a complex path-integral measure and is therefore
not suitable to numerical Monte Carlo simulations. However,
one can consider a purely imaginary source: this strategy
has been developed to study QCD at finite baryon 
density~\cite{imm_baryon_chemical_pot_1,imm_baryon_chemical_pot_2,imm_baryon_chemical_pot_3,imm_baryon_chemical_pot_4},
and has been successfully applied also to the study of 
$\theta$-dependence~\cite{Bhanot:1984rx,Azcoiti:2002vk,theta_immaginario_1,Imachi:2006qq,Aoki:2008gv,first_article_theta_immaginario_SUN,
theta_immaginario_2,DElia:2012pvq,DElia:2013uaf,DElia:2012ifm,
theta_dep_SU3_d_elia}.

In practice, one sets $\theta = -i \theta_I$ and assumes analyticity around $\theta = 0$. The action is modified as follows
\beq
S(\theta) \to S(\theta_I) = S_{\theta = 0} - \theta_I Q \, ,
\eeq
from which it follows that the cumulants of $Q$ are related to the 
corresponding derivatives of the free energy. Using the expression
for $f(\theta)$ in Eq.~(\ref{taylexp}) we have
\begin{equation} \label{imm_theta_fit}
\begin{split}
\frac{k_1(\theta_I)}{V} &= \chi \left[\theta_I -2b_2 \theta_I^3+3b_4 \theta_I^5+O(\theta_I^6)\right] \,, \\
\frac{k_2(\theta_I)}{V} &= \chi \left[1-6b_2 \theta_I^2+15b_4 \theta_I^4+O(\theta_I^5)\right] \, , \\
\frac{k_3(\theta_I)}{V} &= \chi \left[-12b_2 \theta_I+60b_4 \theta_I^3+O(\theta_I^4)\right] \, , \\
\frac{k_4(\theta_I)}{V} &= \chi \left[-12b_2 +180b_4 \theta_I^2+O(\theta_I^3)\right] \, .
\end{split}
\end{equation}
Such equations provide an improved way of
measuring $\chi$ and the $b_{2n}$ coefficients, since one can perform
a global best-fit exploiting the information contained
in the $\theta_I$-dependence of lowest order cumulants,
which is statistically 
more accurate~\cite{theta_dep_SU3_d_elia,DElia:2016jqh}.

In the practical numerical implementation of this procedure,
as in Ref.~\cite{Bonanno:2018xtd}, we have used the geometric
definition $Q_U$ taken after a few 
cooling steps, which is free of renormalizations, to define
the cumulants $k_n$. On the other hand, as explained above, it 
is convenient, for algorithmic reasons, to discretize
the imaginary $\theta$-term in the action by means of 
the non-geometric definition given in Eq.~(\ref{def_non_geo_charge}):
\beq \label{lattice_action_imm_theta}
S_L(\theta_L) = S_L - \theta_L Q_L\, ,
\eeq
so that one would like to know how 
to re-express Eqs.~(\ref{imm_theta_fit})
in terms of $\theta_L$.

As explained above, the relation between $Q_L$ and $Q$ is,
configuration by configuration, $Q_L = Z Q + \eta$, where
$\eta$ is an UV noise.
That means that, for any $n$,
\beq
\frac{d}{d \theta_L} \langle Q^n \rangle &=& 
Z \left( \langle Q^{n+1} \rangle - \langle Q^{n} \rangle \langle Q \rangle \right )
+ (\langle Q^{n}\eta \rangle - \langle Q^{n} \rangle \langle \eta \rangle)
\nonumber \\
&=& Z \left( \langle Q^{n+1} \rangle - \langle Q^{n} \rangle \langle Q \rangle \right)
\eeq
where the second term drops out because $\eta$
is stochastically independent of $Q$.
Based on that, for any $k_n$ one has
\beq
\frac{d k_n}{d \theta_L} = Z \frac{d k_n}{d \theta_I} \, ,
\eeq
so that the Taylor expansion of the cumulants 
in Eq.~(\ref{imm_theta_fit}) can be rewritten
in terms of $\theta_L$ by simply replacing $\theta_I = Z\, \theta_L$,
i.e~the renormalization constant $Z$ represents 
just an additional fit parameter.

\subsection{The Hasenbusch algorithm}

The lattice action in Eq.~(\ref{lattice_action_imm_theta}), being analytic in all fields, can be easily sampled by standard local algorithms. However, these 
algorithms become non-ergodic approaching the continuum limit, 
failing to correctly sample the path-integral. The non-ergodicity is due to the fact that, in the continuum theory, different topological sectors are disconnected and a smooth deformation of the gauge fields cannot change the homotopy class of the configuration. This means that, approaching the continuum limit, 
the number of Monte Carlo steps required to change $Q$ increases 
exponentially with $1/a \sim \xi_L$: this is usually
known as critical slowing down (CSD), and represents a well known 
problem in a wide range of theories sharing the presence of 
topological excitations~\cite{csd_full_QCD_delia, 
deForcrand:1997yw,Lucini:2001ej,DelDebbio:2002xa,Leinweber:2003sj,
critical_slowing_down_review,
obc_paper,
Laio:2015era, Flynn:2015uma,
theta_dep_QCD_N_f_2+1,
Bonati:2017woi}.
Moreover, the problem worsens exponentially increasing 
$N$~\cite{critical_slowing_down_review}, so that
the study of the large-$N$ limit becomes rapidly not feasible, 
even at not-so-large values of $\xi_L$.

Being the CSD related to the existence of non-trivial 
homotopy classes, a possible solution is to switch from
periodic boundary conditions to open boundary
conditions (o.b.c.)~\cite{obc_paper}: topological sectors disappear
and $Q$ can smoothly change between different values.
That does not come for free: finite size effects are more severe, 
constraining to measure observables only in
the bulk of the lattice; $Q$ is no more integer valued
and the information on the $n^{\text{th}}$ order cumulant
is typically obtained in terms of 
integrated $n$-point correlation functions
of the topological charge density, 
with a consequent worsening of the signal-to-noise ratio.

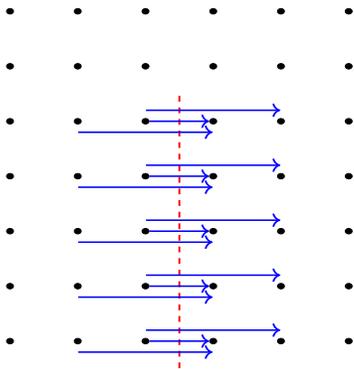
\begin{figure}[!t]
	\centering
	\resizebox{4.8cm}{4.8cm}
	{
	\begin{tikzpicture}
	\node[intermediate point] (9b) at (25mm, 15mm) {};
	
	\node[lattice point] (13) at (0, 20mm) {};
	\node[intermediate point] (13a) at (5mm, 20mm) {};
	\node[intermediate point] (13b) at (5mm, 25mm) {};
	\node[lattice point] (14) at (10mm, 20mm) {};
	\node[intermediate point] (14a) at (15mm, 20mm) {};
	\node[intermediate point] (14b) at (15mm, 25mm) {};
	\node[intermediate point] (14c) at (10mm, 18mm) {};
	\node[intermediate point] (14d) at (10mm, 22mm) {};
	\node[lattice point] (15) at (20mm, 20mm) {};
	\node[intermediate point] (15a) at (25mm, 20mm) {};
	\node[intermediate point] (15b) at (25mm, 25mm) {};
	\node[intermediate point] (15c) at (20mm, 18mm) {};
	\node[intermediate point] (15d) at (20mm, 22mm) {};
	\node[lattice point] (16) at (30mm, 20mm) {};
	\node[intermediate point] (16a) at (35mm, 20mm) {};
	\node[intermediate point] (16b) at (35mm, 25mm) {};
	\node[intermediate point] (16c) at (30mm, 18mm) {};
	\node[intermediate point] (16d) at (30mm, 22mm) {};
	\node[lattice point] (17) at (40mm, 20mm) {};
	\node[intermediate point] (17a) at (45mm, 20mm) {};
	\node[intermediate point] (17b) at (45mm, 25mm) {};
	\node[intermediate point] (17c) at (40mm, 18mm) {};
	\node[intermediate point] (17d) at (40mm, 22mm) {};
	\node[lattice point] (18) at (50mm, 20mm) {};
	
	\node[lattice point] (19) at (0, 30mm) {};
	\node[intermediate point] (19a) at (5mm, 30mm) {};
	\node[intermediate point] (19b) at (5mm, 35mm) {};
	\node[lattice point] (20) at (10mm, 30mm) {};
	\node[intermediate point] (20a) at (15mm, 30mm) {};
	\node[intermediate point] (20b) at (15mm, 35mm) {};
	\node[intermediate point] (20c) at (10mm, 28mm) {};
	\node[intermediate point] (20d) at (10mm, 32mm) {};
	\node[lattice point] (21) at (20mm, 30mm) {};
	\node[intermediate point] (21a) at (25mm, 30mm) {};
	\node[intermediate point] (21b) at (25mm, 35mm) {};
	\node[intermediate point] (21c) at (20mm, 28mm) {};
	\node[intermediate point] (21d) at (20mm, 32mm) {};
	\node[lattice point] (22) at (30mm, 30mm) {};
	\node[intermediate point] (22a) at (35mm, 30mm) {};
	\node[intermediate point] (22b) at (35mm, 35mm) {};
	\node[intermediate point] (22c) at (30mm, 28mm) {};
	\node[intermediate point] (22d) at (30mm, 32mm) {};
	\node[lattice point] (23) at (40mm, 30mm) {};
	\node[intermediate point] (23a) at (45mm, 30mm) {};
	\node[intermediate point] (23b) at (45mm, 35mm) {};
	\node[intermediate point] (23c) at (40mm, 28mm) {};
	\node[intermediate point] (23d) at (40mm, 32mm) {};
	\node[lattice point] (24) at (50mm, 30mm) {};
	
	\node[lattice point] (25) at (0, 40mm) {};
	\node[intermediate point] (25a) at (5mm, 40mm) {};
	\node[intermediate point] (25b) at (5mm, 45mm) {};
	\node[lattice point] (26) at (10mm, 40mm) {};
	\node[intermediate point] (26a) at (15mm, 40mm) {};
	\node[intermediate point] (26b) at (15mm, 45mm) {};
	\node[intermediate point] (26c) at (10mm, 38mm) {};
	\node[intermediate point] (26d) at (10mm, 42mm) {};
	\node[lattice point] (27) at (20mm, 40mm) {};
	\node[intermediate point] (27a) at (25mm, 40mm) {};
	\node[intermediate point] (27b) at (25mm, 45mm) {};
	\node[intermediate point] (27c) at (20mm, 38mm) {};
	\node[intermediate point] (27d) at (20mm, 42mm) {};
	\node[lattice point] (28) at (30mm, 40mm) {};
	\node[intermediate point] (28a) at (35mm, 40mm) {};
	\node[intermediate point] (28b) at (35mm, 45mm) {};
	\node[intermediate point] (28c) at (30mm, 38mm) {};
	\node[intermediate point] (28d) at (30mm, 42mm) {};
	\node[lattice point] (29) at (40mm, 40mm) {};
	\node[intermediate point] (29a) at (45mm, 40mm) {};
	\node[intermediate point] (29b) at (45mm, 45mm) {};
	\node[intermediate point] (29c) at (40mm, 38mm) {};
	\node[intermediate point] (29d) at (40mm, 42mm) {};
	\node[lattice point] (30) at (50mm, 40mm) {};
	
	\node[lattice point] (31) at (0, 50mm) {};
	\node[intermediate point] (31a) at (5mm, 50mm) {};
	\node[intermediate point] (31b) at (5mm, 55mm) {};
	\node[lattice point] (32) at (10mm, 50mm) {};
	\node[intermediate point] (32a) at (15mm, 50mm) {};
	\node[intermediate point] (32b) at (15mm, 55mm) {};
	\node[intermediate point] (32c) at (10mm, 48mm) {};
	\node[intermediate point] (32d) at (10mm, 52mm) {};
	\node[lattice point] (33) at (20mm, 50mm) {};
	\node[intermediate point] (33a) at (25mm, 50mm) {};
	\node[intermediate point] (33b) at (25mm, 55mm) {};
	\node[intermediate point] (33c) at (20mm, 48mm) {};
	\node[intermediate point] (33d) at (20mm, 52mm) {};
	\node[lattice point] (34) at (30mm, 50mm) {};
	\node[intermediate point] (34a) at (35mm, 50mm) {};
	\node[intermediate point] (34b) at (35mm, 55mm) {};
	\node[intermediate point] (34c) at (30mm, 48mm) {};
	\node[intermediate point] (34d) at (30mm, 52mm) {};
	\node[lattice point] (35) at (40mm, 50mm) {};
	\node[intermediate point] (35a) at (45mm, 50mm) {};
	\node[intermediate point] (35b) at (45mm, 55mm) {};
	\node[intermediate point] (35c) at (40mm, 48mm) {};
	\node[intermediate point] (35d) at (40mm, 52mm) {};
	\node[lattice point] (36) at (50mm, 50mm) {};
	
	\node[lattice point] (37) at (0, 60mm) {};
	\node[intermediate point] (37a) at (5mm, 60mm) {};
	\node[intermediate point] (37b) at (5mm, 65mm) {};
	\node[lattice point] (38) at (10mm, 60mm) {};
	\node[intermediate point] (38a) at (15mm, 60mm) {};
	\node[intermediate point] (38b) at (15mm, 65mm) {};
	\node[intermediate point] (38c) at (10mm, 58mm) {};
	\node[intermediate point] (38d) at (10mm, 62mm) {};
	\node[lattice point] (39) at (20mm, 60mm) {};
	\node[intermediate point] (39a) at (25mm, 60mm) {};
	\node[intermediate point] (39b) at (25mm, 65mm) {};
	\node[intermediate point] (39c) at (20mm, 58mm) {};
	\node[intermediate point] (39d) at (20mm, 62mm) {};
	\node[lattice point] (40) at (30mm, 60mm) {};
	\node[intermediate point] (40a) at (35mm, 60mm) {};
	\node[intermediate point] (40b) at (35mm, 65mm) {};
	\node[intermediate point] (40c) at (30mm, 58mm) {};
	\node[intermediate point] (40d) at (30mm, 62mm) {};
	\node[lattice point] (41) at (40mm, 60mm) {};
	\node[intermediate point] (41a) at (45mm, 60mm) {};
	\node[intermediate point] (41b) at (45mm, 65mm) {};
	\node[intermediate point] (41c) at (40mm, 58mm) {};
	\node[intermediate point] (41d) at (40mm, 62mm) {};
	\node[lattice point] (42) at (50mm, 60mm) {};
	
	\node[lattice point] (43) at (0, 70mm) {};
	\node[intermediate point] (43a) at (5mm, 70mm) {};
	\node[intermediate point] (43b) at (5mm, 75mm) {};
	\node[lattice point] (44) at (10mm, 70mm) {};
	\node[intermediate point] (44a) at (15mm, 70mm) {};
	\node[intermediate point] (44b) at (15mm, 75mm) {};
	\node[lattice point] (45) at (20mm, 70mm) {};
	\node[intermediate point] (45a) at (25mm, 70mm) {};
	\node[intermediate point] (45b) at (25mm, 75mm) {};
	\node[lattice point] (46) at (30mm, 70mm) {};
	\node[intermediate point] (46a) at (35mm, 70mm) {};
	\node[intermediate point] (46b) at (35mm, 75mm) {};
	\node[lattice point] (47) at (40mm, 70mm) {};
	\node[intermediate point] (47a) at (45mm, 70mm) {};
	\node[intermediate point] (47b) at (45mm, 75mm) {};
	\node[lattice point] (48) at (50mm, 70mm) {};
	
	\node[lattice point] (49) at (0, 80mm) {};
	\node[intermediate point] (49a) at (5mm, 80mm) {};
	\node[lattice point] (50) at (10mm, 80mm) {};
	\node[intermediate point] (50a) at (15mm, 80mm) {};
	\node[lattice point] (51) at (20mm, 80mm) {};
	\node[intermediate point] (51a) at (25mm, 80mm) {};
	\node[lattice point] (52) at (30mm, 80mm) {};
	\node[intermediate point] (52a) at (35mm, 80mm) {};
	\node[lattice point] (53) at (40mm, 80mm) {};
	\node[intermediate point] (53a) at (45mm, 80mm) {};
	\node[lattice point] (54) at (50mm, 80mm) {};
	
	\draw[red, dashed, thick] (9b) -- (39b);
	
	\draw[blue, thick] [->] (15) -- (16);
	\draw[blue, thick] [->] (15d) -- (17d);
	\draw[blue, thick] [->] (14c) -- (16c);
	\draw[blue, thick] [->] (21) -- (22);
	\draw[blue, thick] [->] (21d) -- (23d);
	\draw[blue, thick] [->] (20c) -- (22c);
	\draw[blue, thick] [->] (27) -- (28);
	\draw[blue, thick] [->] (27d) -- (29d);
	\draw[blue, thick] [->] (26c) -- (28c);
	\draw[blue, thick] [->] (33) -- (34);
	\draw[blue, thick] [->] (33d) -- (35d);
	\draw[blue, thick] [->] (32c) -- (34c);
	\draw[blue, thick] [->] (39) -- (40);
	\draw[blue, thick] [->] (39d) -- (41d);
	\draw[blue, thick] [->] (38c) -- (40c);
	\end{tikzpicture}	
	}
	\caption{The dashed line represents the line defect on the time boundary. Arrows depict links or product of links appearing 
		in the Symanzik action that cross the defect.}
	\label{fig:defect}
\end{figure}
\begin{figure}[!t]
	\centering
	\includegraphics[scale=0.395]{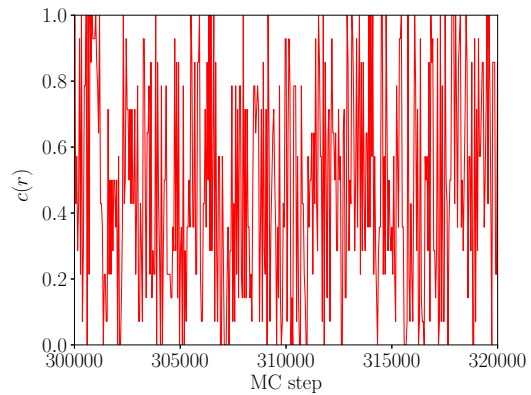}
	\caption{Time evolution of the global parameter 
		$c(r)$ for a given configuration during parallel tempering:
		data refer to $N=51$, $\beta_L=0.6$ and $N_r=15$. 
		One MC step corresponds to 4 sweeps of over-relaxation + 1 sweep of over-heatbath; the showed time window corresponds to $0.0025\%$
		of the total statistics collected for that run.
		In this case swap acceptances range from $60\%$ to $20\%$ 
		and are larger for $c(r)$ closer to 1.}
	\label{fig:replica_swap_evolution}
\end{figure}
\begin{figure}[!h]
	\centering
	\includegraphics[scale=0.395]{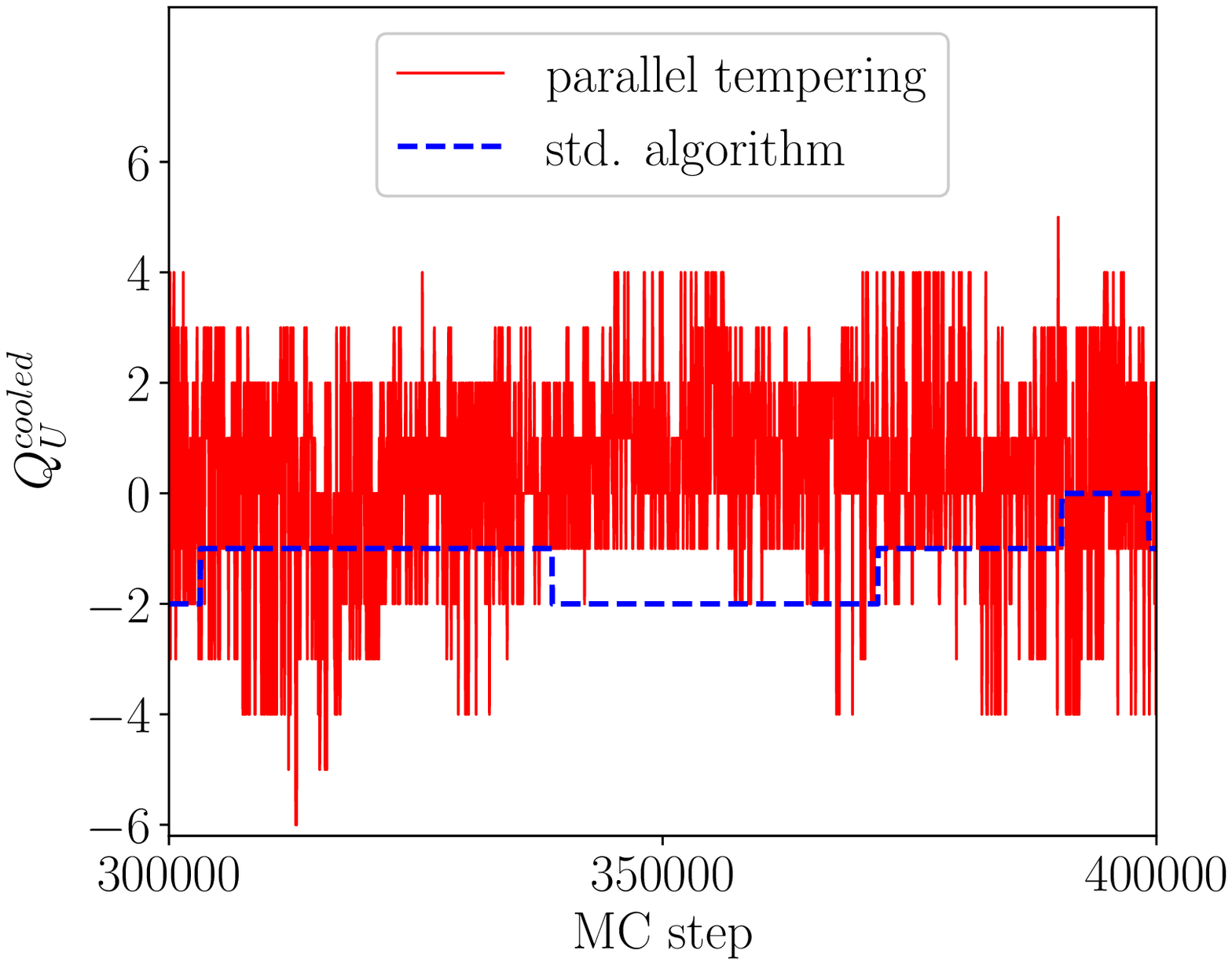}
	\includegraphics[scale=0.395]{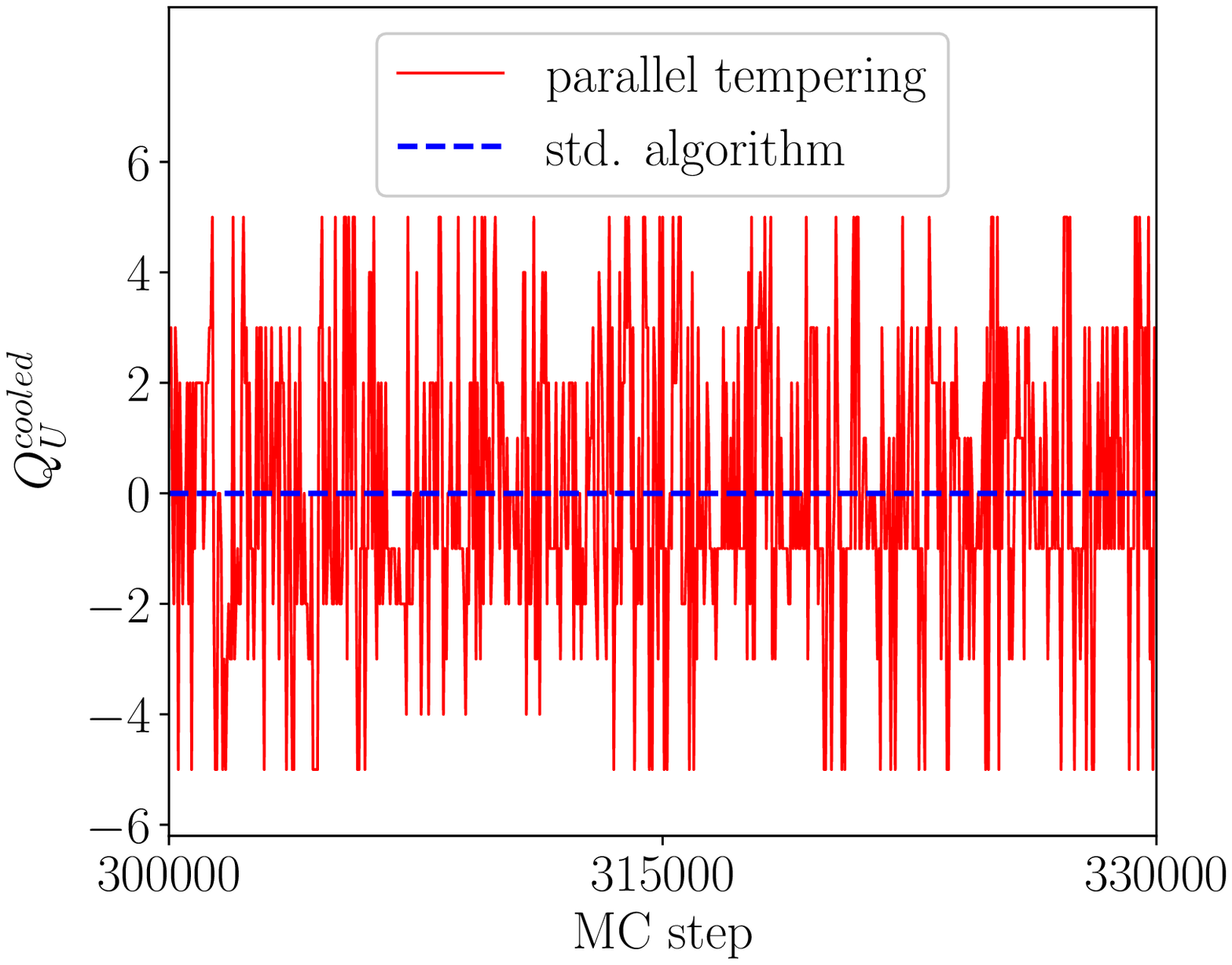}
	\caption{Evolution of $Q_U$ after cooling for $N = 51$ at
		$\beta_L=0.5$ (above) and $\beta_L=0.6$ (below). Results obtained
		by the Hasenbusch algorithm
		are compared to the standard algorithm. The horizontal scale corresponds to the same CPU time for the
		two algorithms and is reported for convenience in MC step
		units (see caption of Fig.~\ref{fig:replica_swap_evolution}). The reported time windows respectively correspond to around $0.013\%$ and $0.0037\%$ of the total statistics collected for the two parallel tempering runs.}
	\label{fig:story_Q_comparison}
\end{figure}

The idea put forward in Ref.~\cite{fighting_slowing_down_cpn}, in order to bypass these complications and still take advantage 
of the improvement of o.b.c., is to 
consider a collection of similar systems, differing among them 
for the value of a global parameter which gradually interpolates between
o.b.c.~and p.b.c.: while each system 
has an independent Monte Carlo evolution, swap of configurations
between different systems are proposed from time to time in a parallel
tempering framework.
In this way, the fast decorrelation of $Q$ achieved for the open system 
is progressively transferred to the periodic one, which is also the 
system where observables are actually measured, thus completely
avoiding the complications of open boundaries.
The analysis of Ref.~\cite{fighting_slowing_down_cpn} shows that it 
suffices to open boundary conditions just along a line defect of length
comparable to $\xi_L$. 
Moreover, hierarchical 
updates around the defect, combined with discrete
translations of the periodic system from time to time, helps
optimizing the algorithm. 
For more details, we refer to Ref.~\cite{fighting_slowing_down_cpn}.

The only differences characterizing our implementation are
that the algorithm is adapted to the Symanzik-improved action, 
and that it is used also for simulations at non-zero imaginary $\theta$.
Actually, a few preliminary tests showed that the choice of
boundary conditions in the $\theta$-term does not affect the 
efficiency of the algorithm: this is expected, since it is 
just the usual action term which develops barriers between
different topological sectors. 
Therefore, in order to avoid further complications, we decided to 
keep periodic boundary conditions in the $\theta$ term 
for all replicas.

The line defect $D$ was put and held fixed on the time boundary: 
$D=\{x \, \vert \, x_0=L-1 \ \wedge \text{ } 0<x_1<L_d\}$, 
where $L_d$ is the defect length: a geometrical representation 
is depicted in Fig.~\ref{fig:defect}. Each link crossing the defect line 
gets multiplied by a factor $c(r)$, where $r$ is the replica index,
$r = 0 , \dots  N_r - 1$.
The interpolation between p.b.c.~
($c(r)=1$) and o.b.c.~($c(r)=0$) 
can be chosen so as to optimize the algorithm, however 
in practice a simple linear interpolation works already well:
$c(r) = 1 - \frac{r}{N_r - 1}$.
The explicit expression of the lattice action 
is the following:
\begin{multline*}
S_L^{(r)} = -2 N \beta_L\sum_{x,\mu}\left\{ k_\mu^{(r)}(x) c_1\Re\left[\bar{U}_\mu(x)\bar{z}(x+\hat{\mu})z(x)\right] \right. \\
\left.
+k_\mu^{(r)}(x+\hat{\mu})k_\mu^{(r)}(x)c_2\Re\left[\bar{U}_\mu(x+\hat{\mu})\bar{U}_\mu(x)\bar{z}(x+2\hat{\mu})z(x)\right]\right\} \, ,
\end{multline*}
where
\beq
k_\mu^{(r)}(x) = 
\begin{cases}
c(r)\, , &\quad x \in D \wedge \mu=0\, ; \\
1\, , &\quad \text{otherwise.}\\
\end{cases}
\eeq
The replica swap was proposed, after every update sweep, 
for each couple of consecutive replicas, and then accepted 
according to a Metropolis step with probability:
\beq
p=\min \left\{1,e^{-\Delta S_L} \right\} \, ,
\eeq
where $\Delta S_L$ is the global change in the action 
of the two involved replicas.
The length of the defect $L_d$ was chosen so that $L_d \sim \xi_L$ at the highest $\beta_L$. 
Concerning $N_r$, for each $N$ we chose it so that, at the highest value of $\beta_L$, the lowest acceptance was not lower than $20\%$.

In Fig.~\ref{fig:replica_swap_evolution} we show how a given
configuration evolves through different values of $c(r)$ in a typical run.
To get the algorithm properly working one should check that swaps happen uniformly, as in the shown example, otherwise the fast decorrelation achieved
for o.b.c.~is not transferred efficiently across the systems.
Finally, in Fig.~\ref{fig:story_Q_comparison} we compare the MC evolution of 
$Q$, with and without using parallel tempering, for 
$N = 51$ and for two different values of $\beta_L$. 
Without parallel tempering the charge is almost frozen, while 
many fluctuations are observed during the same clocktime 
in the other case, allowing us to perform measures which would have been practically impossible with just the local algorithm.

\section{Numerical Results} \label{section_results}

\begin{table}[!t]
	\begin{center}
		\begin{tabular}{ | c || c | c | c | c | c | c | c | c | c | c | } 
			\hline
			& & & & & & & & & &\\[-1em]
			$N$ & $\beta_L$ & $L$ & $\xi_L$ & $\dfrac{L}{\xi_L}$ & $\dfrac{L^2}{\xi_L^2 N}$ & $\theta_{L,\text{\textit{max}}}$ & $N_r$ & $L_d$ &
			\makecell{Stat. \\ $\theta=0$} &
			\makecell{Stat. \\ $\theta\neq0$} \\[0.7em]
			\hline
			\hline
			& & & & & & & & & &\\[-1em]
			\multirow{2}{*}{$21$}
			& 0.68 & 102 & 4.772(4) & 20 & 20 & 0 & 10 & 6 & 76M  & - \\
			& 0.7  & 114 & 5.409(4) & 21 & 21 & 0 & 11 & 6 & 109M & - \\
			\hline
			\hline
			& & & & & & & & & &\\[-1em]
			\multirow{5}{*}{$31$}
			& 0.54 & 56 & 2.218(4) & 25 & 20 & 6 & 10 & 4 & 12M & 10.5M \\
			& 0.56 & 64 & 2.516(4) & 25 & 20 & 6 & 10 & 4 & 12M & 10.5M \\
			& 0.58 & 72 & 2.856(5) & 25 & 20 & 6 & 10 & 4 & 12M & 10.5M \\
			& 0.6  & 82 & 3.239(5) & 25 & 20 & 6 & 10 & 4 & 16M & 21.7M \\
			& 0.62 & 92 & 3.672(6) & 25 & 20 & 6 & 10 & 4 & 15M & 27.3M \\
			\hline
			\hline
			& & & & & & & & & &\\[-1em]
			\multirow{7}{*}{$41$}
			& 0.51 & 58  & 1.952(3) & 29 & 21 & 6   & 13 & 4 & 19M & 11.2M \\
			& 0.53 & 64  & 2.213(4) & 29 & 20 & 6   & 13 & 4 & 17M & 11.9M \\
			& 0.55 & 74  & 2.517(4) & 29 & 21 & 6   & 13 & 4 & 16M & 15.4M \\
			& 0.57 & 82  & 2.840(5) & 29 & 20 & 6   & 13 & 4 & 20M & 21M   \\
			& 0.59 & 92  & 3.226(5) & 28 & 20 & 6   & 13 & 4 & 22M & 20.3M \\
			& 0.61 & 104 & 3.655(7) & 28 & 20 & 6.5 & 13 & 4 & 17M & 31.9M \\
			& 0.65 & 132 & 4.698(6) & 28 & 20 & 0   & 15 & 5 & 39M & -    \\
			\hline
			\hline
			& & & & & & & & & &\\[-1em]
			\multirow{6}{*}{$51$}
			& 0.5  & 62  & 1.902(3) & 33 & 21 & 6    & 15 & 4 & 17M & 11.2M \\
			& 0.52 & 70  & 2.104(4) & 33 & 22 & 6    & 15 & 4 & 19M & 11.2M \\
			& 0.54 & 78  & 2.445(4) & 32 & 20 & 6    & 15 & 4 & 19M & 11.9M \\
			& 0.56 & 88  & 2.779(5) & 32 & 20 & 6    & 15 & 4 & 18M & 11.9M \\
			& 0.58 & 100 & 3.153(6) & 32 & 20 & 6.5  & 15 & 4 & 17M & 29.7M \\
			& 0.6  & 114 & 3.560(7) & 32 & 20 & 6.5  & 15 & 4 & 18M & 28.6M \\
			\hline
		\end{tabular}
	\end{center}
	\caption{Summary of the simulation parameters adopted for all values of $N$. We also report the total accumulated statistics, where each measure was taken after every parallel tempering step. The imaginary-$\theta$ fit was always performed in the range $[0,6]$ with 7 points in steps of $\delta \theta_L= 1$, except for the $\beta_L$ with $\theta_{L,\text{\textit{max}}}=6.5$ where 11 points were taken ($\delta \theta_L=1$ up to $\theta_L=3$ and then $\delta \theta_L=0.5$); $\theta_{L,\text{\textit{max}}}=0$ indicates that no simulation at imaginary $\theta$ has been performed: in this case a single high-statistics run at $\theta = 0$ has been used to determine $\chi$ and $\xi$ at the same time. $N_r$ indicates the number of replicas used for parallel tempering, the number of hierarchical levels was always 3 (see \cite{fighting_slowing_down_cpn} for more details on the hierarchical update).}
	\label{table_info_simulations}
\end{table}

In Tab.~\ref{table_info_simulations} we summarize
the parameters and statistics of the simulations 
performed in the present study, which regarded 
$N = 21,\, 31, \, 41$ and $51$. In addition, we will also 
consider results obtained at lower $N$ in previous studies,
in order to investigate the large-$N$ behavior of the theory. Results for $N = 21$ and $N = 31$ have been already reported
in Ref.~\cite{Bonanno:2018xtd}, but using lower statistics
and/or smaller correlation lengths than in the present work.

Statistics at $\theta_L = 0$ are generally larger
because in this case, apart from the cumulants of the cooled 
charge $Q_U$, we determined the value of the correlation
length $\xi_L$, which is needed with the best possible 
precision since it affects the final precision on 
the continuum extrapolation of $\xi^2 \chi$.
Statistical errors on the cumulants have been estimated 
by means of a bootstrap analysis.

In the following we will first discuss the impact 
of finite size effects, in order to justify the choices
for the lattice sizes used in our simulations. 
Then we will illustrate 
the procedure and discuss the systematics both for 
the analytic continuation from imaginary $\theta$ and for 
the extrapolation to the continuum limit of our results;
in this respect, we notice that some of our results for 
$\xi^2 \chi$ have been produced without relying on analytic continuation
(see Tab.~\ref{table_info_simulations}), in order to 
provide a robust test that the systematics of analytic continuation
and continuum extrapolation are actually under control.

Finally, we will study the large-$N$ expansion of $\xi^2 \chi$, $b_2$, $b_4$ based on our and on previous numerical results,
illustrating the systematics, the comparison with
existing analytic predictions and the estimate of further terms 
in the expansion.

\subsection{Finite size effects}

In all of our simulations, and for each fixed
value of $N$, we decided to approach the continuum limit
keeping the ratio $L/\xi_L$ fixed, so as to ensure
that the physical volume is kept fixed.
In general, taking a ratio $L/\xi_L \gg 1$ should
ensure that finite size effects be negligible.
However, as discussed in Ref.~\cite{Aguado:2010ex},
this is not the case for the $\theta$-dependence of $2d$ $CP^{N-1}$ models in the large-$N$ limit:
the large-$N$ limit and the thermodynamic limit do not commute
for this class of models, and wrong results might be obtained if the former limit is taken first, leading in practice to the more 
restrictive condition $(L/\xi_L)^2\gg N$.
Since this is strictly related to the particular
large-$N$ behavior of the $\theta$-dependence of the theory,
we will try to give an intuitive explanation of this condition (see also Ref.~\cite{Bolognesi:2019rwq} for a related discussion).

Basically, the reason can be traced back to the fact that both $\xi^2 \chi$ and the $b_{2n}$ coefficients 
vanish in the large-$N$ limit; indeed,
$\xi^2 \chi$ vanishes as $1/N$, thus, indicating 
by $l = L a $ the size in physical units,
one expects for large $N$
\beq
\langle Q^2 \rangle = \chi\, l^2 
\simeq  \frac{l^2}{2 \pi \xi^2 N} \, .
\eeq
If $l/\xi=L/\xi_L$ is kept fixed while $N \to \infty$, one ends
up with a system where $\langle Q^2 \rangle \ll 1$: in these conditions, the distribution of $Q$ 
is strongly peaked at $Q = 0$, with only rare occurences of 
$Q = \pm 1$ and even rarer occurences of higher values of $|Q|$.
It is easy to check that such a distribution leads 
to $b_{2n}$ coefficients which are very close to those predicted by 
DIGA, i.e., for instance,
$b_2  = -\langle Q^4 \rangle_c / (12 \langle Q^2 \rangle) \simeq -1/12$.
Let us stress that
this happens in practice in many other situations, like for instance
for 4$d$ Yang-Mills theories in the high-$T$ phase,
where $\chi$ vanishes rapidly above $T_c$ and one
cannot afford to take very large lattice volumes,
ending up again with $\langle Q^2 \rangle \ll 1$;
however, in that case the system is really close to DIGA
even in the thermodynamical limit, so that, luckily enough, 
this does not imply significant systematic errors
(see Ref.~\cite{todaro} for a thorough discussion about this point).

\begin{figure}[!t]
	\centering
	\includegraphics[scale=0.4]{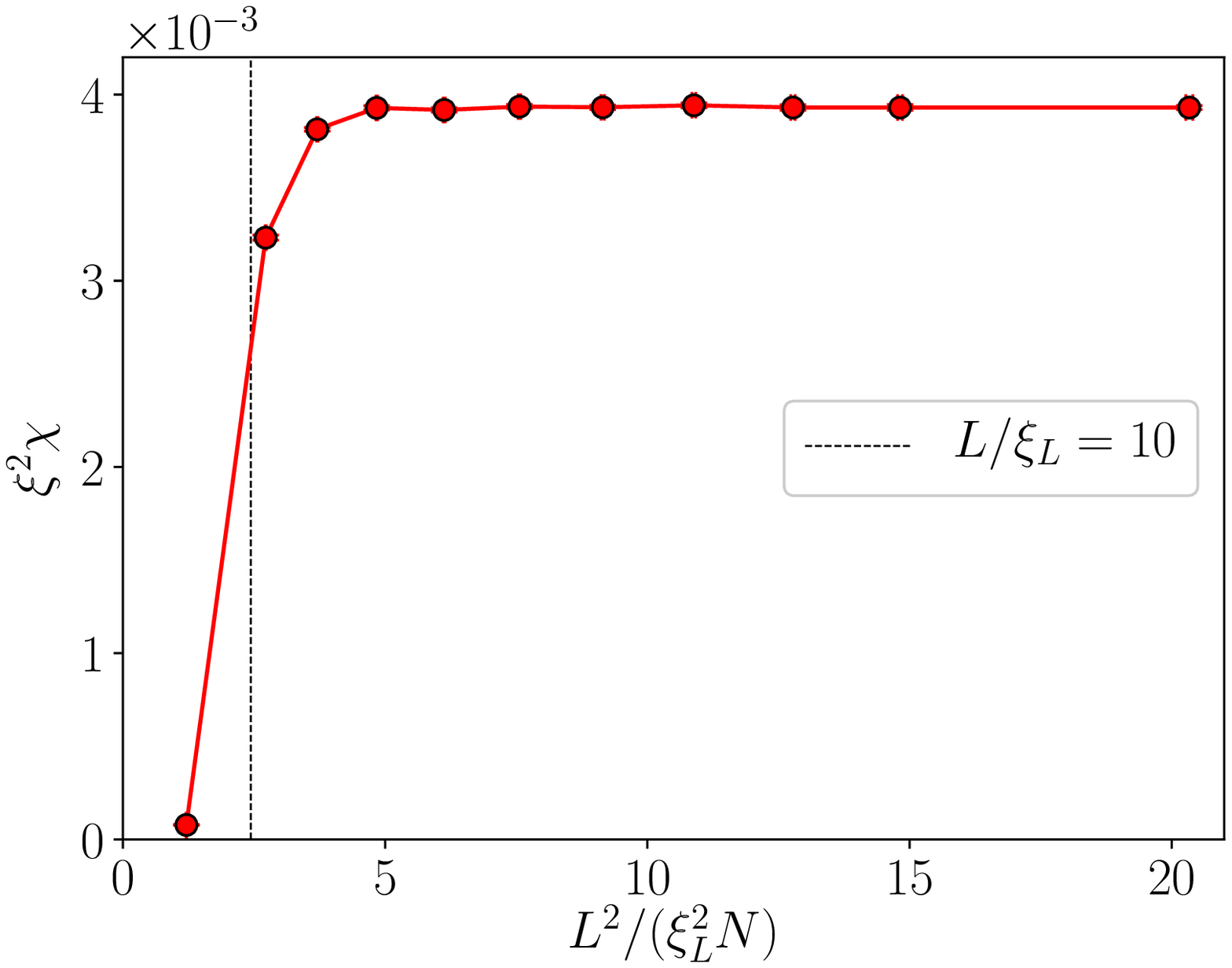}
	\hspace*{-0.3cm}
	\includegraphics[scale=0.4]{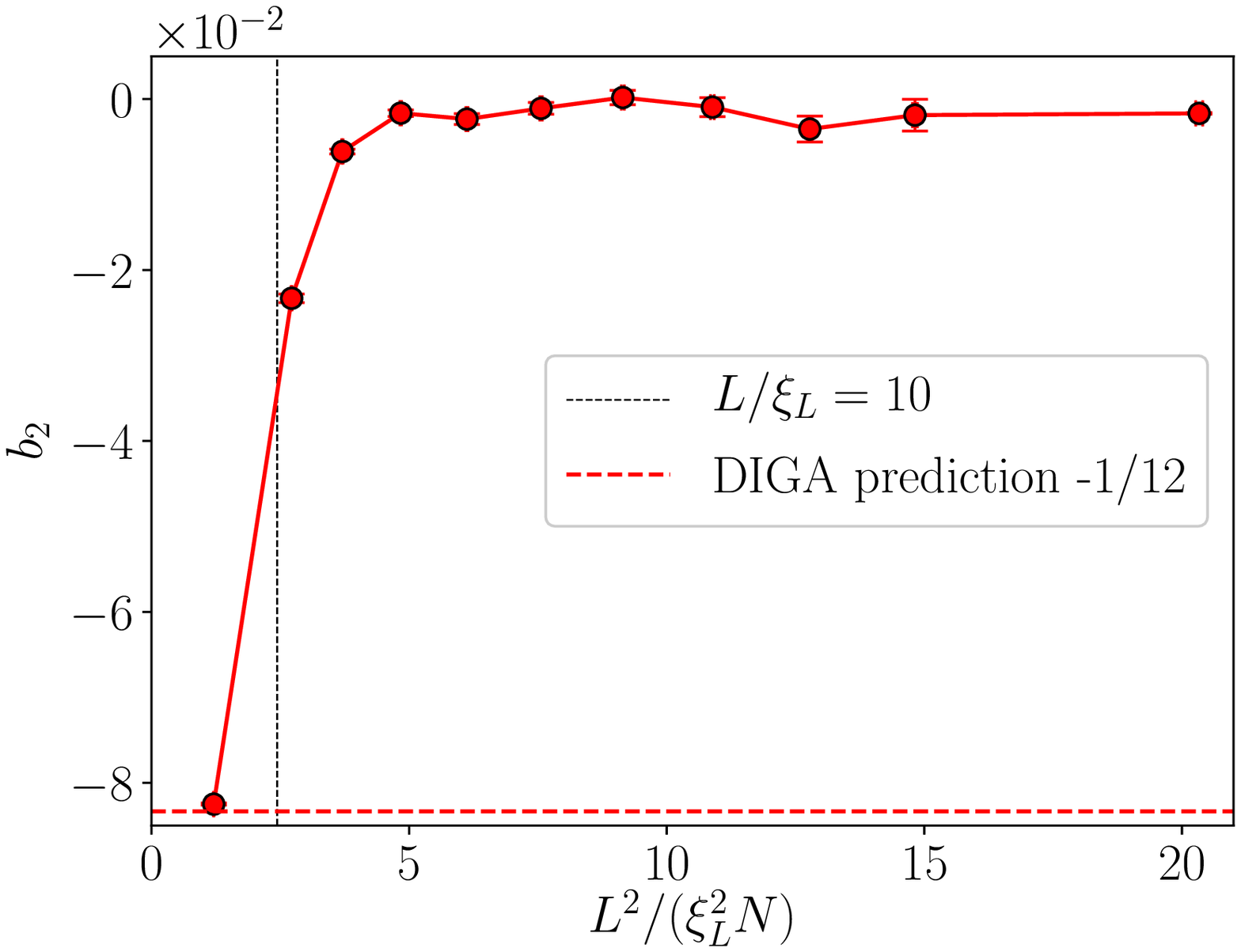}
	\caption{Study of the dependence of $\xi^2 \chi$ and $b_2$ on the lattice size for $N=41$ and $\beta_L=0.57$. Lattice sizes range from $L=20$ to $L=82$. Results were obtained by parallel tempering and using 
		the same setup (apart from the lattice size) 
		reported in Tab.~\ref{table_info_simulations}. The value employed for $\xi_L$ is, for all sizes, the one obtained for the largest $L$.}
	\label{fig:finite_size_effects}
\end{figure}

For 2$d$ $CP^{N-1}$ models, instead, the large-$N$ limit
is qualitatively different from DIGA,
indeed one expects $b_2 \propto 1/N^2$, which is quite different
from DIGA predictions. Stated in another way, the 
particularity of 2$d$ $CP^{N-1}$ models is that while 
global topological excitations become rarer and rarer
as $N \to \infty$, they never become really dilute, 
as one would naively expect.
The additional condition
$(L/\xi_L)^2\gg N$ ensures that $\langle Q^2 \rangle$ is 
at least of $O(1)$ or larger, so that the non-trivial 
interactions between topological excitations, which characterize
the large-$N$ limit, can be properly taken into account.

\begin{table*}[!t]
	\begin{center}
		\begin{tabular}{|c|c|c||c|c|c|c|c|c|c|}
			\hline
			& & & & & & & & &\\[-1em]
			$N$ & $\beta_L$ & $\xi_L$ & $Z$ & $a^2 \chi\cdot 10^3$ & $\xi^2 \chi\cdot 10^3$ & $b_2\cdot 10^3$ & $b_4\cdot 10^6$ & $\tilde{\chi}^2$ & dof \\
			\hline
			\hline
			& & & & & & & & &\\[-1em]
			\multirow{2}{*}{$21$}
			& 0.68 & 4.772(4) & -        & 0.3404(6)  & 7.750(18) & -         & -         & -   & -\\
			& 0.7  & 5.409(4) & -        & 0.2647(4)  & 7.746(16) & -         & -         & -   & -\\
			\hline
			\hline
			& & & & & & & & &\\[-1em]
			\multirow{5}{*}{$31$}
			& 0.54 & 2.218(4) & 0.876(2) & 1.0907(19) & 5.365(21) & -2.90(9)  & -4.6(1.5) & 1.0 & 17 \\
			& 0.56 & 2.517(4) & 0.892(2) & 0.8353(17) & 5.290(22) & -2.67(10) & -1.7(1.6) & 0.8 & 17 \\
			& 0.58 & 2.856(5) & 0.895(2) & 0.6471(17) & 5.279(24) & -2.75(12) & -2.8(1.9) & 0.7 & 17 \\
			& 0.6  & 3.240(5) & 0.903(2) & 0.4993(13) & 5.240(21) & -2.82(11) & -4.3(1.7) & 0.9 & 17 \\
			& 0.62 & 3.672(6) & 0.913(3) & 0.3876(12) & 5.227(23) & -2.59(11) & -2.1(1.7) & 1.0 & 17 \\
			\hline
			\hline
			& & & & & & & & &\\[-1em]
			\multirow{7}{*}{$41$}
			& 0.51 & 1.953(3) & 0.896(2) & 1.0642(16) & 4.057(14) & -1.99(8) & -3.5(1.3) & 0.9 & 17 \\
			& 0.53 & 2.213(4) & 0.903(2) & 0.8222(15) & 4.027(15) & -1.85(8) & -2.1(1.3) & 1.3 & 17 \\
			& 0.55 & 2.517(4) & 0.915(2) & 0.6295(14) & 3.987(17) & -1.84(9) & -2.2(1.4) & 0.6 & 17 \\
			& 0.57 & 2.840(5) & 0.924(2) & 0.4873(11) & 3.930(16) & -1.69(9) & -1.1(1.4) & 1.2 & 17 \\
			& 0.59 & 3.226(5) & 0.928(3) & 0.3775(11) & 3.930(17) & -1.72(12)& -1.0(1.8) & 1.3 & 17 \\
			& 0.61 & 3.655(7) & 0.928(3) & 0.2936(10) & 3.922(20) & -1.76(10)& -2.1(1.3) & 1.3 & 29 \\
			& 0.65 & 4.698(6) & -        & 0.1772(7)  & 3.912(18) & -        & -         & -   & - \\
			\hline
			\hline
			& & & & & & & & &\\[-1em]
			\multirow{6}{*}{$51$}
			& 0.50 & 1.902(3) & 0.917(2) & 0.8970(16) & 3.244(13) & -1.38(8) & -2.2(1.3) & 1.2 & 17 \\
			& 0.52 & 2.104(4) & 0.919(2) & 0.7297(16) & 3.231(14) & -1.55(10)& -4.4(1.5) & 1.2 & 17 \\
			& 0.54 & 2.445(4) & 0.929(2) & 0.5317(13) & 3.178(13) & -1.39(11)& -3.4(1.7) & 0.9 & 17 \\
			& 0.56 & 2.779(5) & 0.932(3) & 0.4125(13) & 3.186(16) & -1.23(12)& -0.4(2.0) & 0.8 & 17 \\
			& 0.58 & 3.153(6) & 0.943(3) & 0.3172(10) & 3.154(16) & -1.27(9) & -1.7(1.1) & 1.2 & 29 \\
			& 0.60 & 3.560(7) & 0.938(4) & 0.2478(10) & 3.140(17) & -1.52(18)& -4.2(1.5) & 1.7 & 29 \\
			\hline
		\end{tabular}
	\end{center}
	\caption{Summary of results obtained for $\xi^2 \chi$, $b_2$, $b_4$, $Z$ and $\xi_L$ for all explored values of $\beta_L$ and $N$. All results have been obtained through the imaginary-$\theta$ fit of the first 3 cumulants, except for the two measures at $N=21$ and for the measure at the highest $\beta_L$ at $N=41$, where no fit was performed. For all fits we also report the number of dof and the reduced $\tilde{\chi}^2$.}
	\label{summary_results}
\end{table*}

In order to illustrate the above considerations in practice,
in Fig.~\ref{fig:finite_size_effects} we report the behavior
of $\xi^2\chi$ and $b_2$ 
as a function of the lattice size for $N = 41$.
It clearly appears that in the small volume limit 
the system is described by DIGA ($b_2 \simeq -1/12$), 
and that finite size corrections are still significant
for $L/\xi_L = 10$.
We stress, however, that the range of 
$L/\xi_L$ for which finite size effects are visible 
at this particular value of $N$ is still compatible 
with the range observed for other generic (non topological)
observables in the same class of models, 
i.e.~$L/\xi_L \lesssim 20$~\cite{Rossi:1993nz}: in order to really observe discrepancies with respect 
to the indications of Ref.~\cite{Rossi:1993nz} one
should study a case for which $L / \xi_L \gtrsim 20$ 
and $L^2 / (N \xi_L^2) \lesssim 1$, however that requires
$N$ to be of the order of a few hundreds.

In any case, in our simulations we kept $L^2/(\xi_L^2 N) \sim 20$
for all explored values of 
$N$, meaning that $L/\xi_L$ depends on $N$
(see Tab.~\ref{table_info_simulations}). We stress that 
the condition $L^2/(\xi_L^2 N) \gg 1$ was also 
ensured in the numerical simulations reported
in Ref.~\cite{Bonanno:2018xtd}.

\subsection{Analytic continuation from imaginary $\theta$ and continuum extrapolation}

\begin{figure}
\includegraphics[scale=0.4]{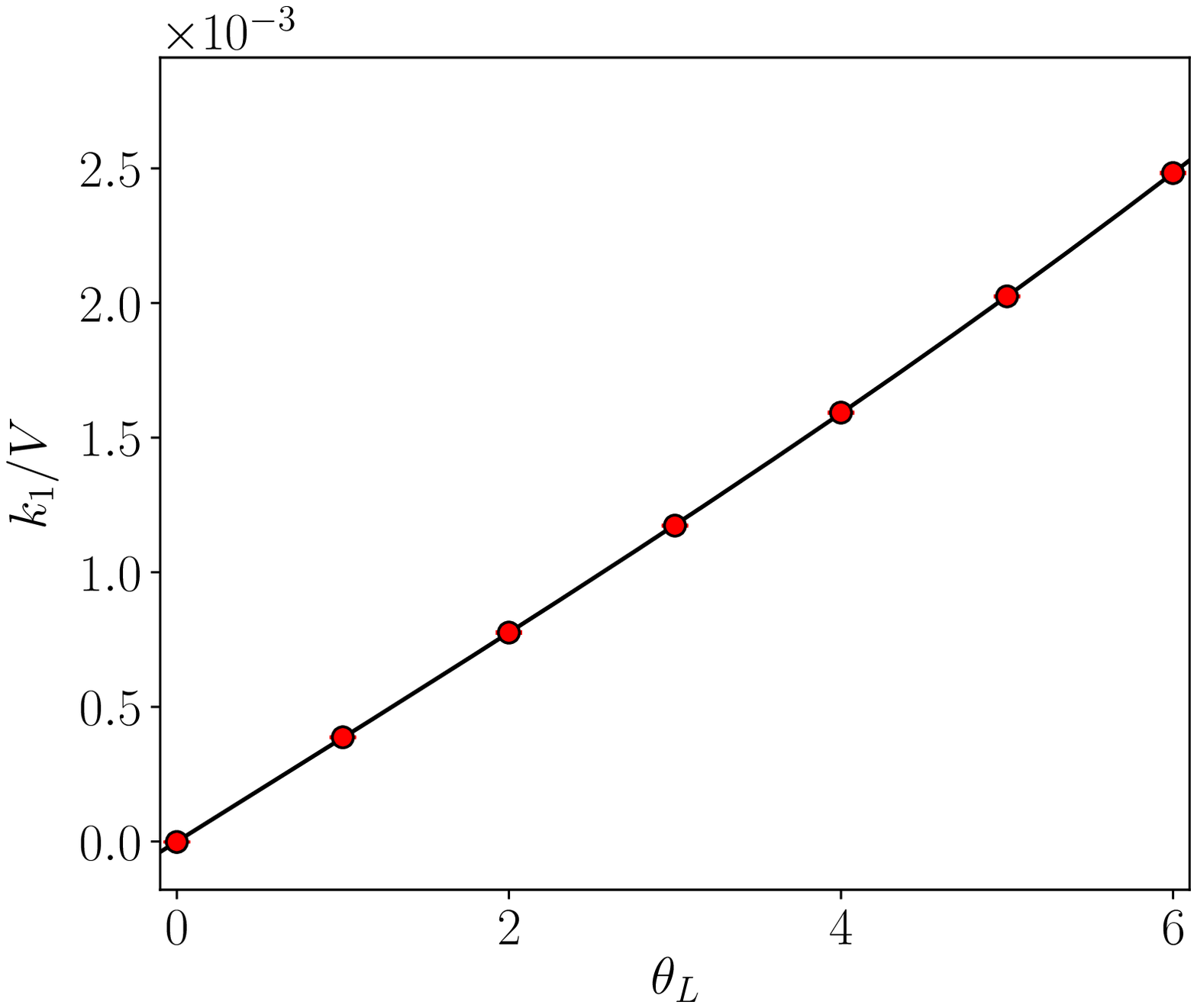}
\includegraphics[scale=0.4]{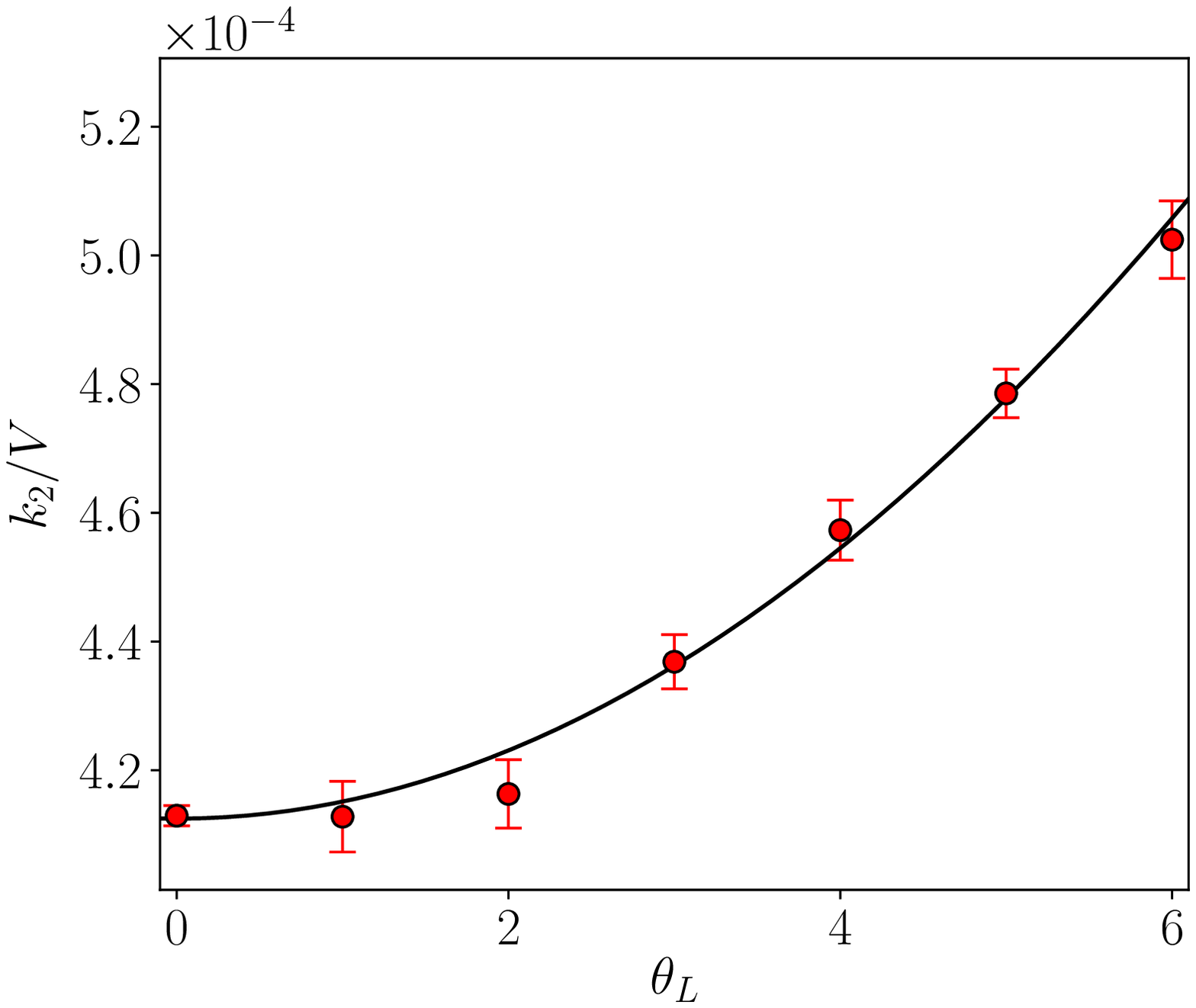}
\includegraphics[scale=0.4]{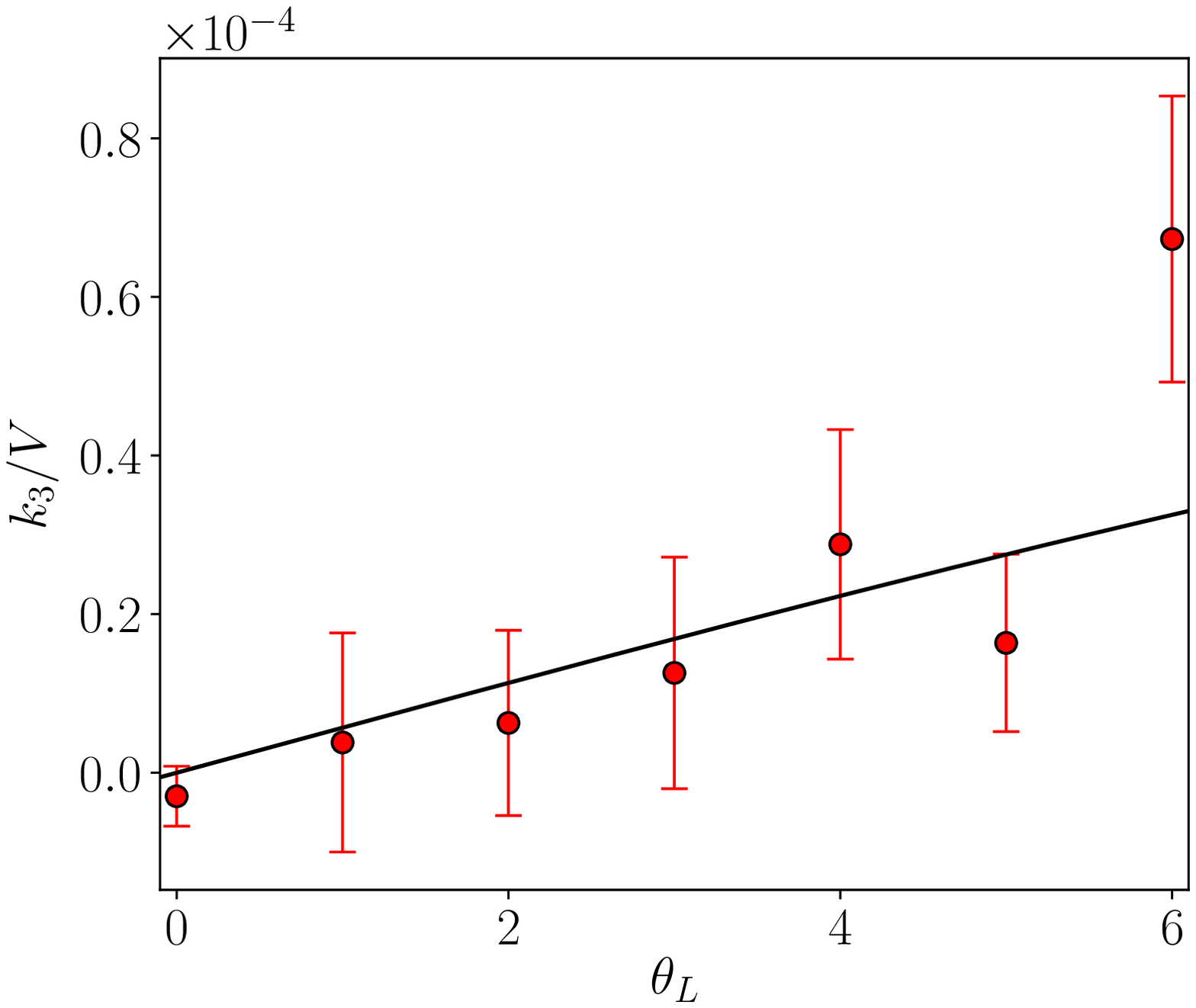}
\caption{Example of the global imaginary-$\theta$ fit for $N=51$ and $\beta_L=0.56$, exploiting the 3 lowest order cumulants and fitting up to $O(\theta_L^6)$ terms. The best fit yields a reduced $\tilde{\chi}^2=0.83$ with 17 degrees of freedom (dof).}
\label{fig:imm_theta_fit_example}
\end{figure}

In Fig.~\ref{fig:imm_theta_fit_example} we show 
an example of the global imaginary-$\theta$ fit
discussed in Section~\ref{subsec_imm_theta}.
The best-fit procedure was performed 
according to Eq.~(\ref{imm_theta_fit}) and 
for just the first 3 cumulants in all cases,
exploiting the whole available imaginary $\theta$ range.
In order to assess the impact of systematic
effects related to analytic continuation, 
we have tried in each case to change the range
fit and the order (i.e.~the truncation) of the fit polynomial, verifying that the variation of the
fit parameters was within
statistical errors, or adding it to the total error otherwise.

A complete summary of the results obtained for 
$Z$, $\xi_L$, $\xi^2 \chi$, $b_2$ and $b_4$ at all explored values
of $N$ and $\beta$ is reported in Tab.~\ref{summary_results}.

\begin{figure}[!htb]
\includegraphics[scale=0.4]{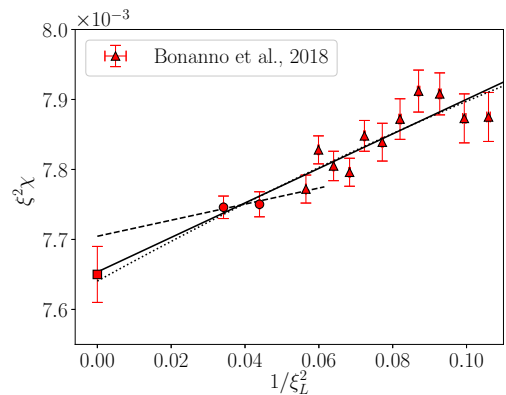}
\caption{Continuum extrapolation of $\xi^2 \chi$ for $N=21$. The solid line represents a linear fit in $1/\xi_L^2$ in the whole range, the dashed line a linear fit on a restricted range and the dotted one a quadratic fit in $1/\xi_L^2$ in the whole range. Round points represent lattice measures from this study, triangle points are taken from Ref.~\cite{Bonanno:2018xtd}, while the square point is the continuum extrapolated value. Best fit results and the corresponding reduced $\tilde{\chi}^2$ values are reported in Tab.~\ref{table_continuum_limit_systematics}.}.
\label{fig:cont_limit_example_0}
\end{figure}

Results collected at different values of $\beta_L$ have then 
been used to obtain continuum extrapolated quantities. In order
to discuss the procedure that we have adopted in all cases,
we illustrate in details our analysis for the continuum
extrapolation of $\xi^2 \chi$ at 
$N = 21$. 
Results at finite $\xi_L$ 
are reported in Fig.~\ref{fig:cont_limit_example_0} as a function
of $1/\xi_L^2$, and
include both results from Ref.~\cite{Bonanno:2018xtd} (obtained via analytic 
continuation) and from this 
work (obtained just from simulations at $\theta = 0$). In general, for the lattice discretization adopted in this work, one expects 
corrections to the continuum limit for a generic observable $O$, including only the two lowest non-trivial terms, to be as follows:
\beq
\braket{O}_L(\xi_L) = \braket{O}_{\text{\textit{cont}}} + \frac{A}{\xi_L^2} + 
\frac{B}{\xi_L^4} \, .
\eeq
In general, for all the quantities considered in this study,
a linear extrapolation in $1/\xi_L^2$, i.e.~setting $B = 0$, has worked
perfectly well, i.e.~with reduced $\tilde \chi^2$ of order 1 for the best 
fit, in the whole explored range of $\xi_L$. However, in order to correctly 
assess the 
impact of systematic errors related to the extrapolation, 
we have also analyzed the effect of including 
the non-linear term ($B \neq 0$), or of considering the linear fit in 
a restricted range of $\xi_L$, i.e.~discarding points which are farther from 
the continuum limit.
The systematics of this procedure are reported 
in Tab.~\ref{table_continuum_limit_systematics}, some of the best
fits are reported in Fig.~\ref{fig:cont_limit_example_0} as well.
After considering the observed systematics, our final determination
for $\xi^2 \chi$ at $N = 21$ has been 
$\xi^2 \chi = 0.00765(4)$, to be compared with 
$\xi^2 \chi = 0.00759(5)$ from Ref.~\cite{Bonanno:2018xtd}, 
$\xi^2 \chi = 0.00767(5)$ from Ref.~\cite{fighting_slowing_down_cpn} (adopting
a different discretization) and 
$\xi^2 \chi = 0.00800(20)$ from Ref.~\cite{critical_slowing_down_review}.

\begin{figure}[!htb]
\centering
\includegraphics[scale=0.4]{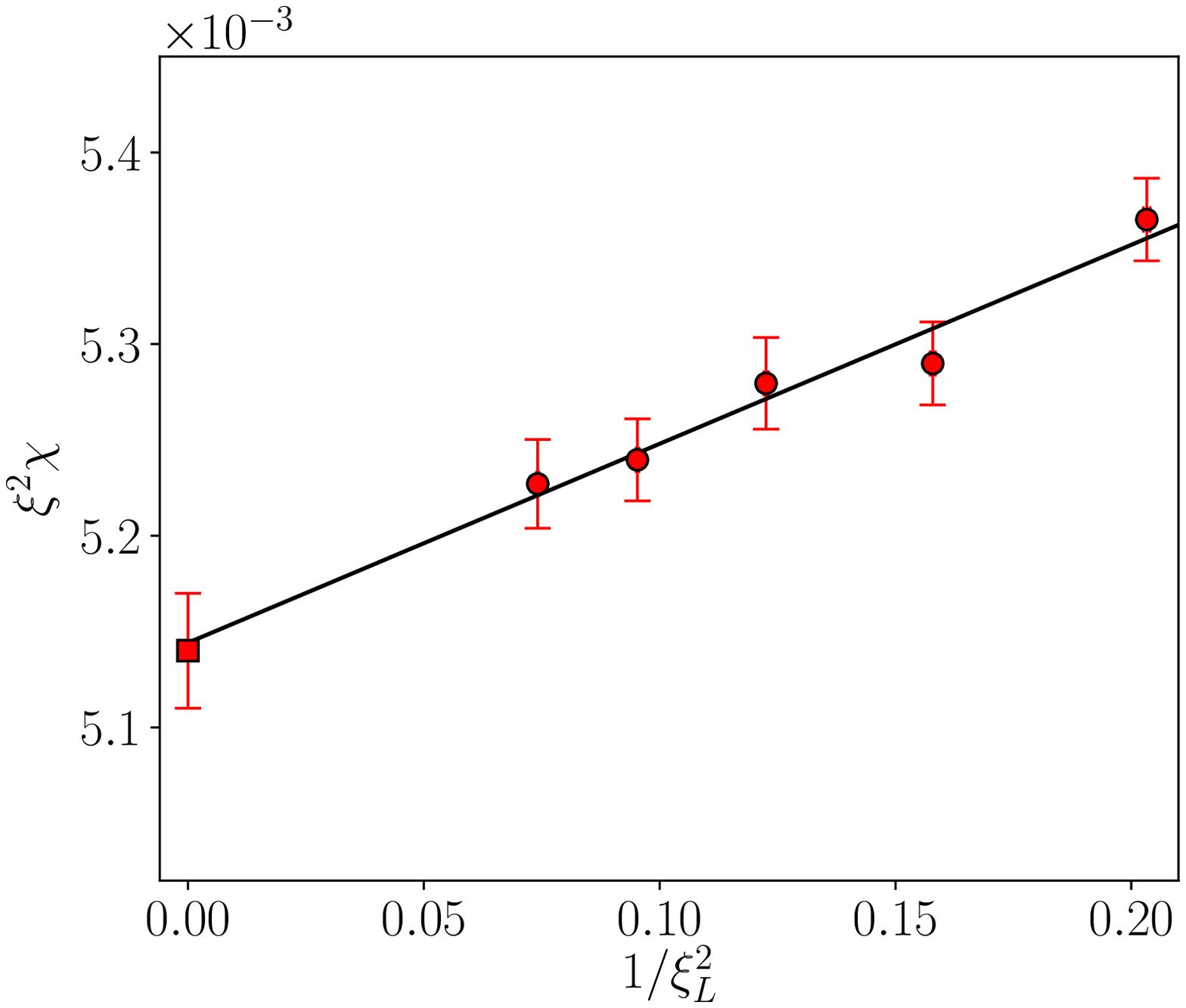}
\includegraphics[scale=0.4]{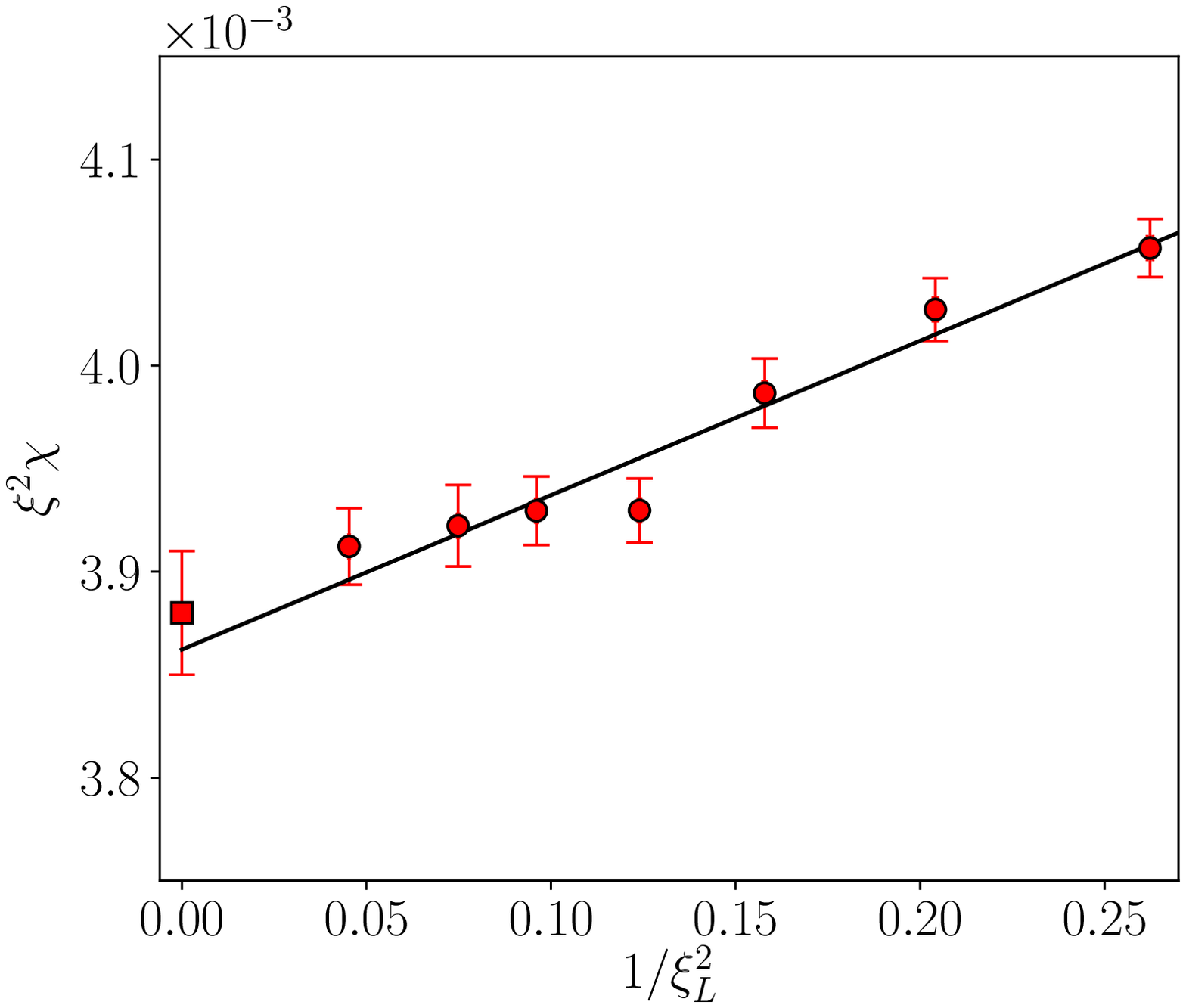}
\includegraphics[scale=0.4]{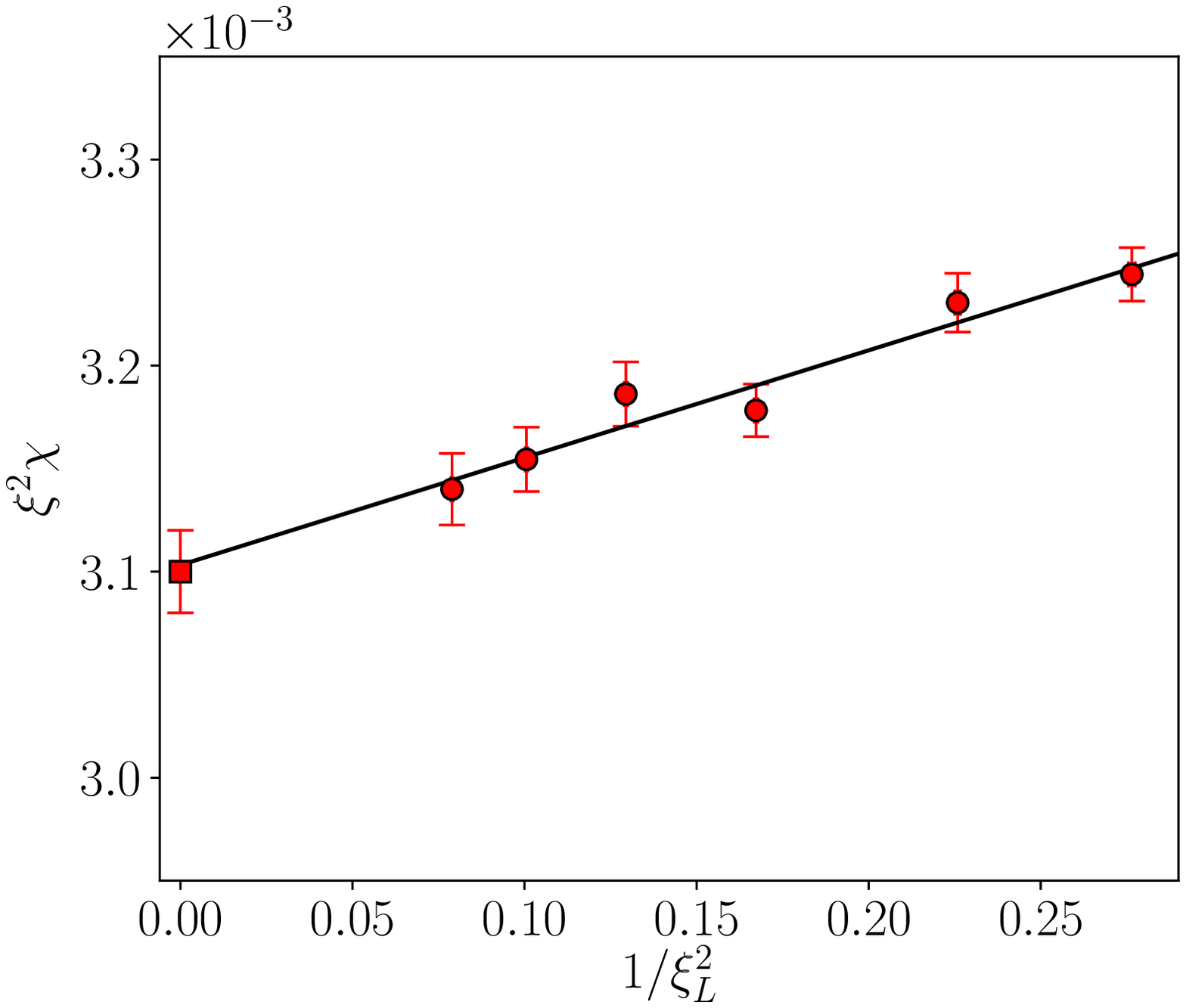}
\caption{From top to bottom: continuum extrapolations of $\xi^2 \chi$ for 
	$N=31,\,41$ and $51$. The solid line represents a linear fit in $1/\xi_L^2$ in the whole range, all best fits yield a reduced $\chi^2$ of order 1.
	Systematics of the continuum extrapolation 
	have been estimated as for $N = 21$, the final extrapolated 
	result which is plotted for $1/\xi_L^2 = 0$ includes such
	systematics.}
\label{fig:cont_limit_example_1}
\end{figure}
\begin{figure}[!htb]
\centering
\includegraphics[scale=0.4]{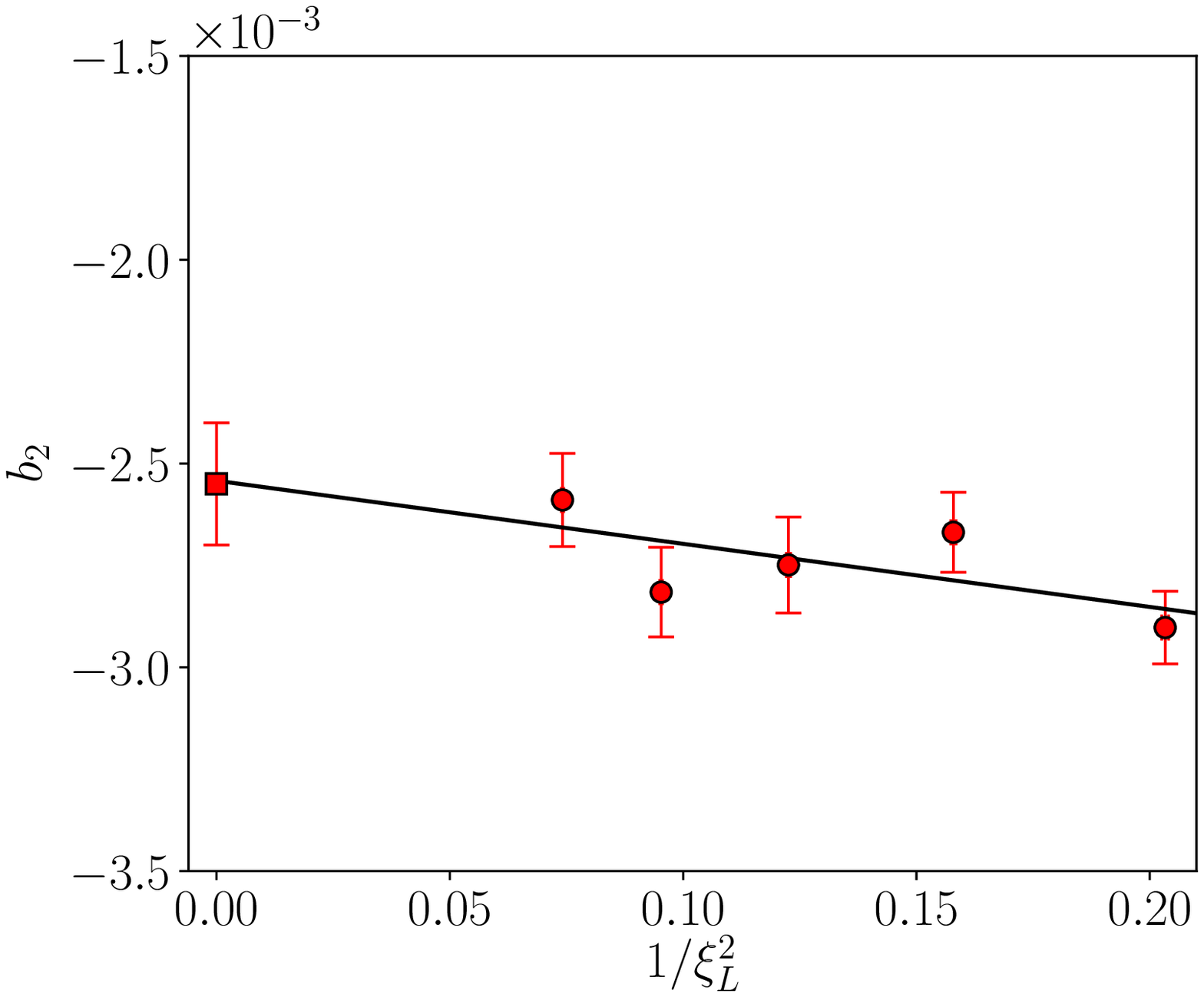}
\includegraphics[scale=0.4]{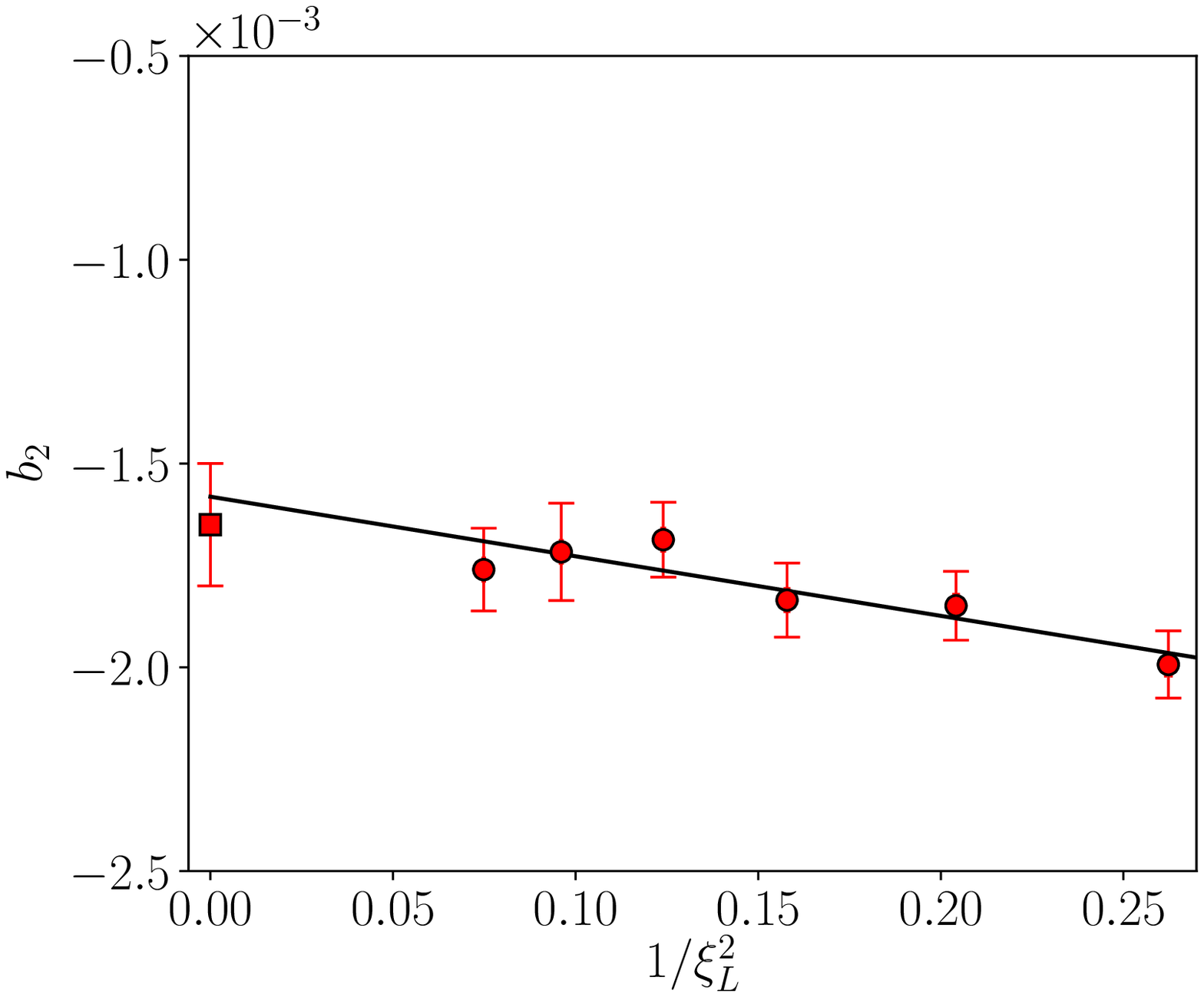}
\hspace*{0.05cm}
\includegraphics[scale=0.4]{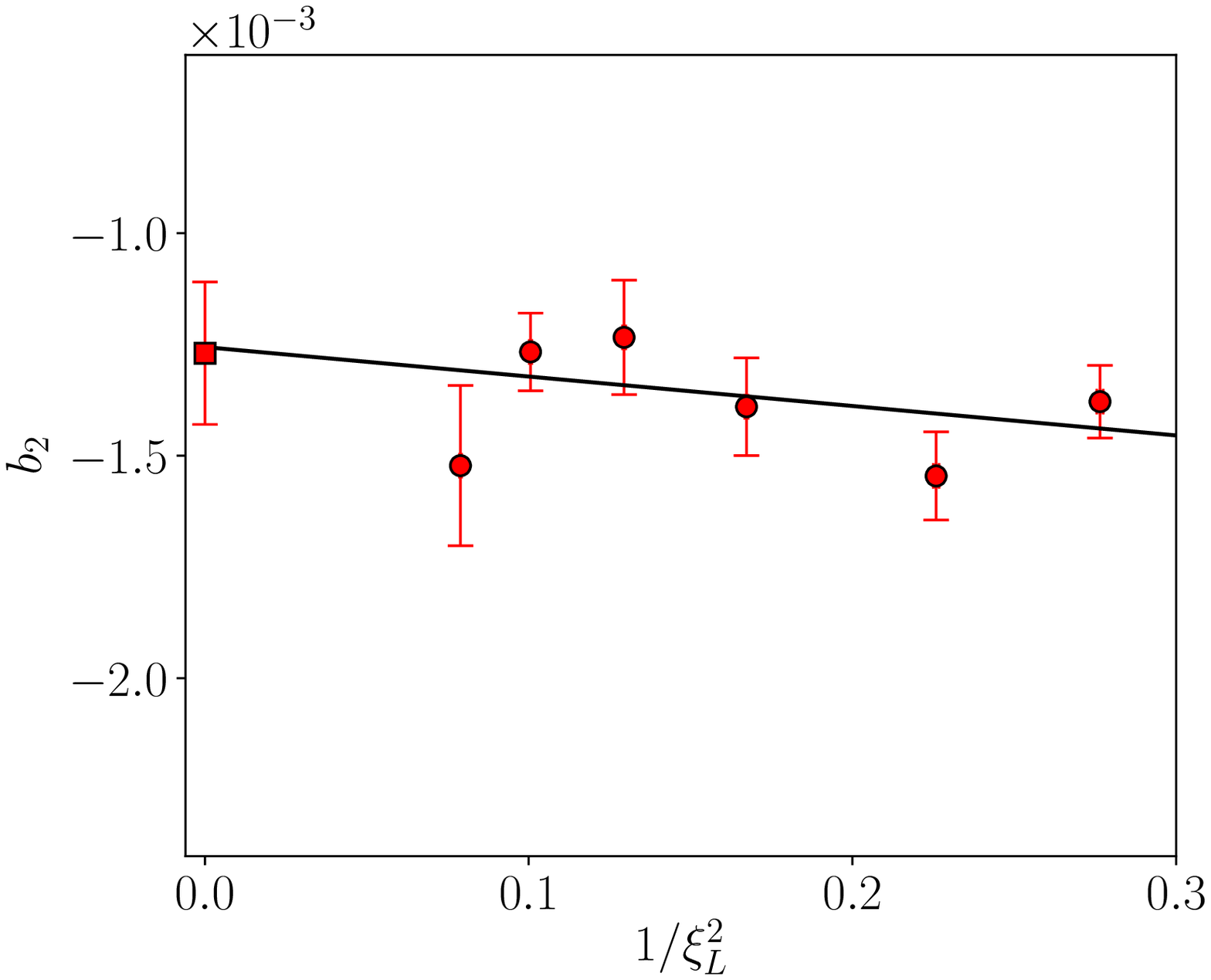}
\caption{From top to bottom: continuum extrapolations of $b_2$ for 
	$N=31,\,41$ and $51$. The solid line represents a linear fit in $1/\xi_L^2$ in the whole range, all best fits yield a reduced $\chi^2$ of order 1.
	Systematics of the continuum extrapolation 
	have been estimated as for $N = 21$, the final extrapolated 
	result which is plotted for $1/\xi_L^2 = 0$ includes such
	systematics.}
\label{fig:cont_limit_example_2}
\end{figure}
\begin{figure}[!htb]
\centering
\includegraphics[scale=0.4]{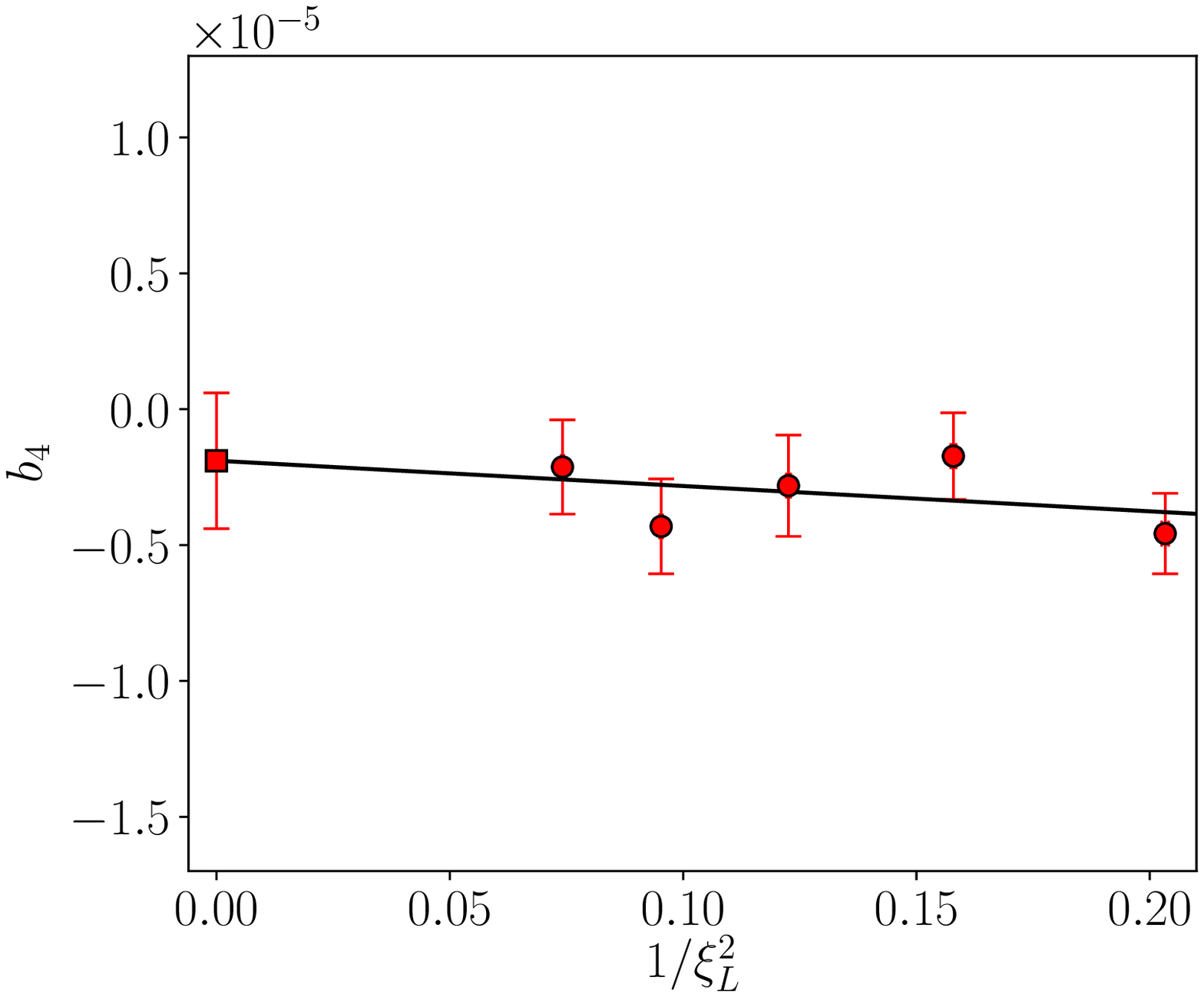}
\includegraphics[scale=0.4]{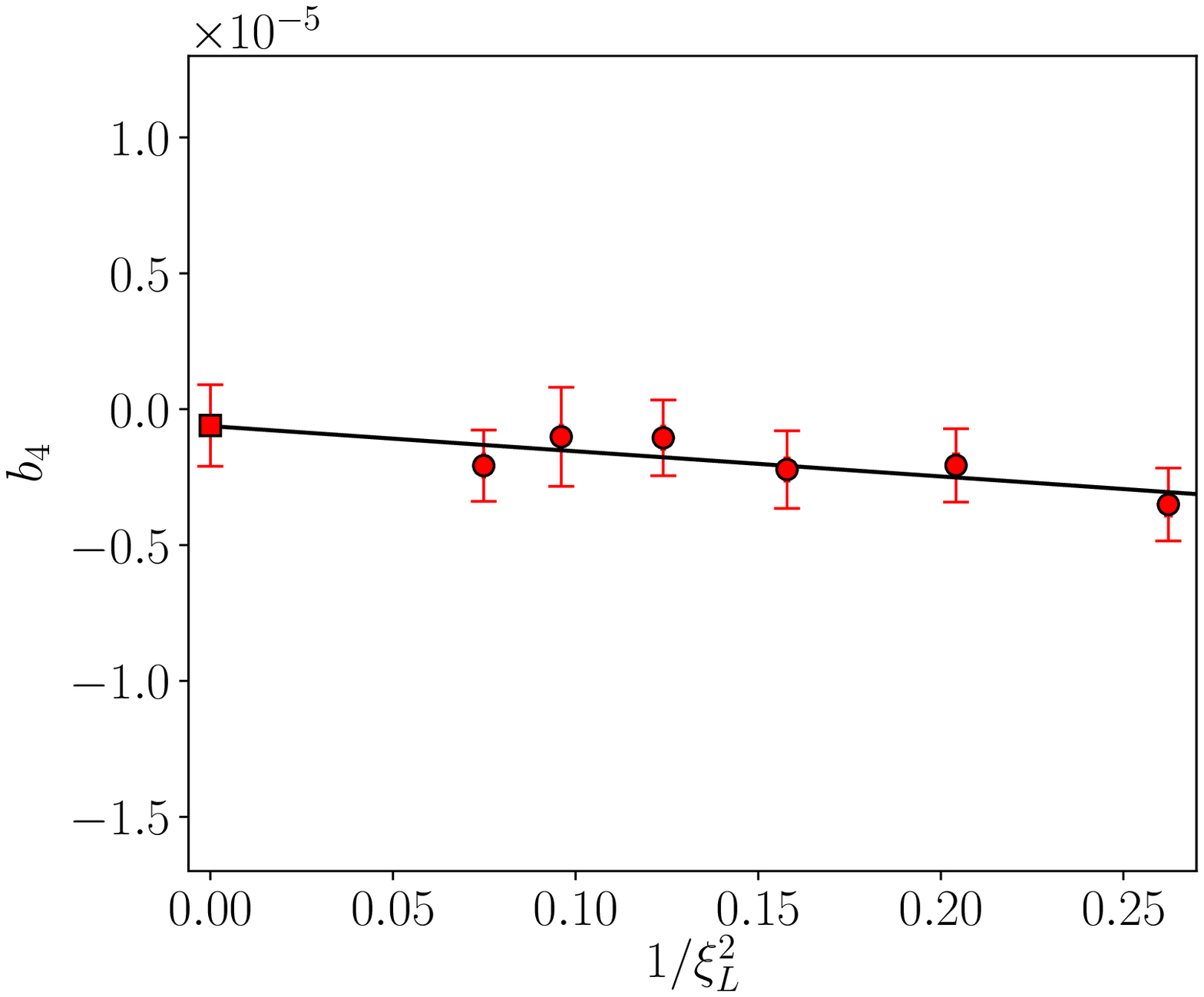}
\hspace*{0.1cm}
\includegraphics[scale=0.4]{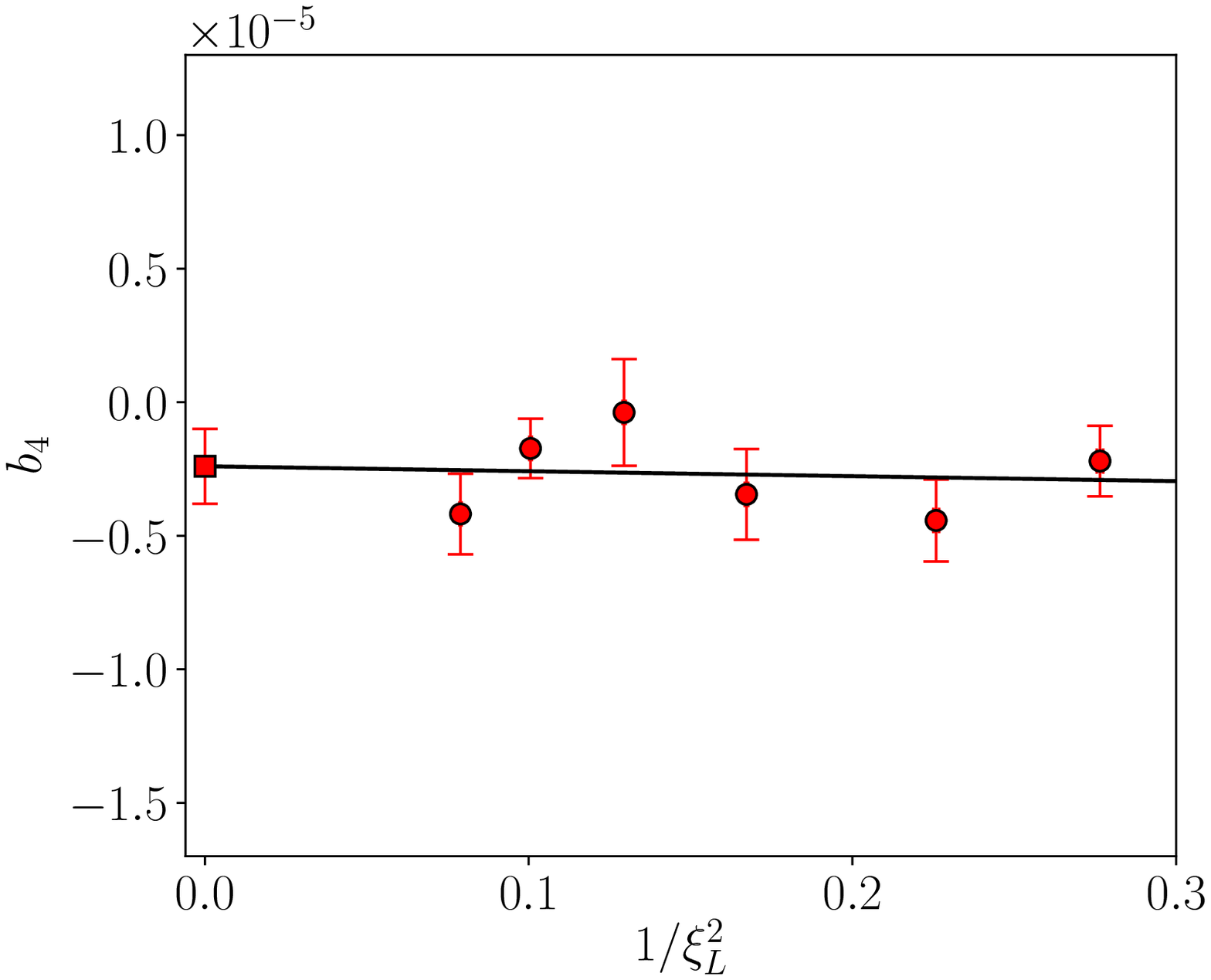}
\caption{From top to bottom: continuum extrapolations of $b_4$ for $N=31,\,41$ and $51$. The solid line represents a linear fit in $1/\xi_L^2$ in the whole range, all best fits yield a reduced $\chi^2$ of order 1.
	Systematics of the continuum extrapolation 
	have been estimated as for $N = 21$, the final extrapolated 
	result which is plotted for $1/\xi_L^2 = 0$ includes such
	systematics.}
\label{fig:cont_limit_example_3}
\end{figure}
\begin{table}[!htb]
\begin{center}
\begin{tabular}{ | c || c | c | c | } 
\hline
& & & \\[-1em]
$\xi_{L,\text{\textit{min}}}$ included & $\xi^2 \chi \cdot 10^3$ & $\tilde{\chi}^2$ & dof \\
\hline
\hline
& & & \\[-1em]
3.07 (quadratic fit) & 7.64(6)   & 1.09 & 10 \\
\hline
& & & \\[-1em]
3.07 (linear fit)    & 7.654(21) & 0.99 & 11 \\
3.28                 & 7.636(24) & 0.91 & 9  \\
3.49                 & 7.652(27) & 0.88 & 7  \\
3.72                 & 7.66(3)   & 1.14 & 5  \\
3.95                 & 7.66(4)   & 1.21 & 3  \\
4.21                 & 7.71(5)   & 0.10 & 1  \\
\hline
\end{tabular}
\end{center}
\caption{Summary of systematics for the continuum extrapolation of $\xi^2 \chi$ for $N=21$. Our final determination in this case is 
$\xi^2 \chi = 0.00765(4)$.}
\label{table_continuum_limit_systematics}
\end{table}

The procedure above has been repeated for all explored quantities
and for all $N$. In 
Figs.~\ref{fig:cont_limit_example_1}, 
\ref{fig:cont_limit_example_2} and \ref{fig:cont_limit_example_3} 
we show the 
continuum extrapolations for 
$\xi^2 \chi$, $b_2$ and $b_4$ for $N = 31,41$ and 51,
for simplicity of figure reading we report just the linear
best fits over the whole range, even if the final continuum
determinations take into account the whole systematics,
as in the example above.

Final results are shown in Tab.~\ref{table_large_N_measures}, where we report the continuum limit of $\xi^2 \chi$, $b_2$ and $b_4$ for $N=21,\, 31,\, 41$ and $51$, along with continuum results of these observables for other values of $N$ taken from Refs.~\cite{Bonanno:2018xtd, fighting_slowing_down_cpn}.
\begin{table}[!htb]
\begin{center}
\begin{tabular}{ | c || c | c | c | } 
\hline
& & & \\[-1em]
$N$ & $\xi^2\chi \cdot 10^3$ & $b_2 \cdot 10^3$ & $b_4 \cdot 10^5$ \\
\hline
& & & \\[-1em]
9  & 20.00(15)  & -13.90(13) & 2.04(18)  \\
10 & 17.37(8)   &     -      &     -     \\
11 & 15.24(12)  & -10.7(4)   & 2.3(5)    \\
13 & 12.62(9)   & -9.1(3)    & -0.6(5)   \\
15 & 10.87(11)  & -6.7(3)    & 0.5(6)    \\
21 & 7.65(4)    & -5.0(5)    & -1.7(1.2) \\
26 & 6.14(5)    & -3.0(4)    & -0.3(5)   \\
31 & 5.14(3)    & -2.55(15)  & -0.2(3)   \\
41 & 3.88(3)    & -1.65(15)  & -0.06(15) \\
51 & 3.10(2)    & -1.27(16)  & -0.24(14) \\
\hline
\end{tabular}
\end{center}
\caption{Summary of continuum determinations of $\xi^2 \chi$, $b_2$ and $b_4$ for several values of $N$. Results for $N=9,\, 11,\,13,\,15$ and $26$, as well as $b_2$ and $b_4$ for $N=21$, are taken from Ref.~\cite{Bonanno:2018xtd} while the one for $N=10$ from Ref.~\cite{fighting_slowing_down_cpn}.
The result obtained for $\xi^2 \chi$ at $N = 41$ is in good agreement
(less than 1$\sigma$) with that obtained using a different
discretization in Ref.~\cite{fighting_slowing_down_cpn}
($\xi^2 \chi = 0.00391(2)$), where however no determination was given
for $b_2$ and $b_4$; for $N = 31$ instead one should compare 
with the results of Ref.~\cite{Bonanno:2018xtd}
($\xi^2 \chi = 0.00503(6)$, $b_2 = -0.00231(22)$), also in this 
case the agreement is reasonable (1.6$\sigma$ and 0.9$\sigma$ respectively).}
\label{table_large_N_measures}
\end{table}

\subsection{Results for $\xi^2 \chi$ and its large-$N$ scaling}

\begin{figure}[!htb]
\centering
\includegraphics[scale=0.4]{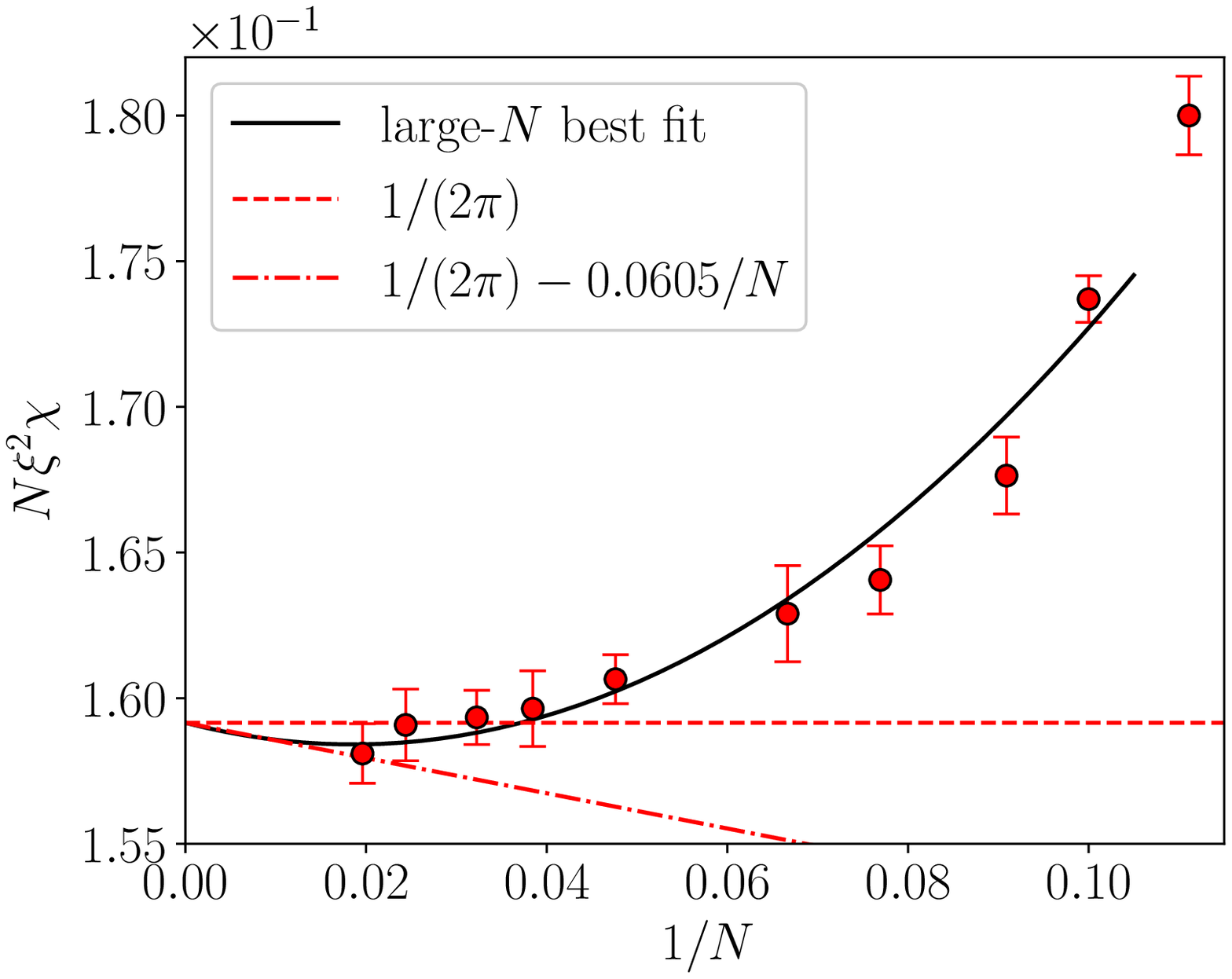}
\caption{Behavior of $N\xi^2 \chi$ as a function of $1/N$ compared to LO (dashed line) and NLO (dot-dashed line) 
analytic computations of the $1/N$ series. 
The solid line is the result of a best fit to data with 
$N \geq 10$
where the LO is fixed to the analytic result and 
the NLO and NNLO are fitted (see Tab.~\ref{table_large_N_chi_vs_1_on_N} for the complete systematics).}
\label{fig_grande_N_susc_top}
\end{figure}

The main purpose of this section is to make use of the results 
reported in Tab.~\ref{table_large_N_measures} to investigate the 
large-$N$ behavior of the topological susceptibility and compare it with 
analytical predictions. In Fig.~\ref{fig_grande_N_susc_top} we plot the quantity
$N \xi^2 \chi$, which should approach a constant for $N \to \infty$,
as a function of $1/N$, together with the LO and NLO analytic 
computations, and one of our best fits to be discussed in the following.

\begin{figure}[!htb]
\centering
\includegraphics[scale=0.4]{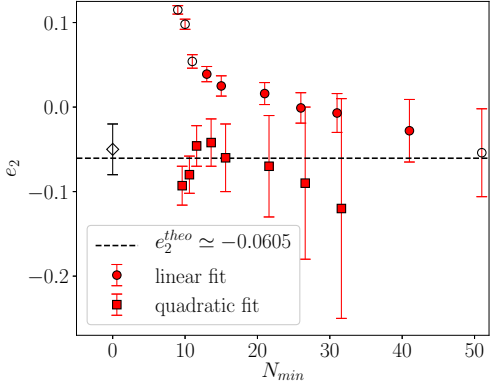}
\caption{Summary of the fit systematics for the NLO coefficient $e_2$ of the $1/N$ expansion of $\xi^2 \chi$. Empty round points depict determinations of $e_2$ from fits with a reduced $\chi^2$ with $p$-value smaller than $5\%$ or non existent. The empty diamond point represents our overall estimation: $e_2=-0.05(3)$.}
\label{fig_e2}
\end{figure}

Similarly to what has been done in 
Ref.~\cite{Bonanno:2018xtd}, we fit our data with a function of the type
\beq\label{fit_function_large_N_xi2_chi}
N \xi^2 \chi = e_1 + \frac{e_2}{N} + \frac{e_3}{N^2}
+ \frac{e_4}{N^3}
\eeq
which includes up to N$^3$LO corrections in the $1/N$ expansion:
all best-fit systematics are summarized 
in Tab.~\ref{table_large_N_chi_vs_1_on_N};
in all cases, the LO term has been fixed to the 
well established analytic prediction
$e_1 = 1 / (2 \pi)$. The systematics for the $e_2$ coefficient
are also plotted in Fig.~\ref{fig_e2} in order 
to make the discussion clearer.

If one sets $e_3 = e_4 = 0$, thus allowing only for the NLO correction, 
acceptable or marginally acceptable best fits are obtained when 
data for $N < 13$ are discarded. Nevertheless, results obtained for
$e_2$ are not stable and show a systematic drift as the fit 
range is changed, suggesting that NNLO corrections could be important. 
They are positive and in clear disagreement 
with the analytic predictions $e_2 = -0.0605$~\cite{calcolo_e_2}
(as also reported in previous literature) 
if the fitted range is large enough, but decrease systematically
and out-of-the-errors 
as the fit range is restricted to larger and larger values of 
$N$, finally becoming negative compatible with the analytic predictions,
even if within very large error bars (see Fig.~\ref{fig_e2}).

On the other hand, when the NNLO correction is included in the fit,
$e_3 \neq 0$, results obtained for $e_2$ are reasonably stable 
and compatible with the analytic 
predictions.
Moreover, the NNLO term $e_3$ appears quite stable as well, also when the NLO 
coefficient $e_2$ is fixed to the theoretically predicted value, and even when a further term ($e_4 \neq 0$) is included in the fit,
so that we can provide a quite conservative estimate
$e_3=1.5(5)$, which considers all variations observed for this 
coefficient in the various fits. The values obtained for $e_4$ are not sufficiently precise 
or stable to allow for any estimate, however we can state it is 
of $O(10)$.

Finally, we point out that the values obtained for the reduced $\tilde{\chi}^2$ for the fits including N$^2$LO and N$^3$LO corrections are generally low: one possible interpretation is that we have been too conservative in estimating the errors on continuum extrapolated quantities, that might also explain	why the fit including just $e_2$ yields acceptable values of $\tilde \chi^2$ but is not stable.

Present results clarify why previous lattice studies observed 
a positive deviations with respect to the LO prediction, in contradiction
with the fact that $e_2$ is negative: the NNLO term
has an opposite sign and, given its estimated magnitude, it is 
expected to dominate until $N > |e_3/e_2| \sim 20-30$. Our analysis fully supports the analytic prediction of $e_2$.
On the other hand, if we had to provide an independent determination of $e_2$,
a conservative estimate based on our systematics would 
be $e_2 = -0.05(3)$, which is still quite inaccurate
despite the large numerical effort; the difficulty is clearly related to 
the fact that $e_2$ turns out to be quite small in magnitude, both 
with respect to $e_1$ and $e_3$, so that it is hardly detectable.

\begin{table}[!htb]
\begin{center}
\begin{tabular}{ | c | c | c | c | c | c | c | } 
\hline
& & & & & &\\[-1em]
$N_{\text{\textit{min}}}$ & $e_1$ & $e_2$ & $e_3$ & $e_4$ & $\tilde \chi^2$ & dof\\
\hline
\hline
& & & & & &\\[-1em]
51 & $1 / (2 \pi)$ & -0.054(52)    & & & - & 0 \\
41 & "             & -0.028(37)    & & & 0.49 & 1 \\
31 & "             & -0.007(23)    & & & 0.51 & 2 \\
26 & "             & -0.001(18)    & & & 0.42 & 3 \\
21 & "             &  0.016(13)    & & & 0.71 & 4 \\
15 & "             &  0.025(12)    & & & 1.03 & 5 \\
13 & "             &  0.039(9)     & & & 1.6 & 6 \\
11 & "             &  0.054(8)     & & & 2.8  & 7 \\
10 & "             &  0.098(6)     & & & 11   & 8 \\
9  & "             &  0.115(5)     & & & 15   & 9 \\
\hline
\hline
& & & & & &\\[-1em]
31 & $1 / (2 \pi)$ & -0.12(13)  & 3.9(4.3) & & 0.21 & 1 \\
26 & "             & -0.09(9)   & 2.8(2.7) & & 0.16 & 2 \\
21 & "             & -0.07(6)   & 2.2(1.4) & & 0.13 & 3 \\
15 & "             & -0.06(4)   & 1.8(8)   & & 0.13 & 4 \\
13 & "             & -0.042(28) & 1.4(5)   & & 0.16 & 5 \\
11 & "             & -0.046(24) & 1.5(4)   & & 0.15 & 6 \\
10 & "             & -0.080(22) & 2.2(3)   & & 1.02 & 7 \\
9  & "             & -0.093(23) & 2.4(3)   & & 1.35 & 8 \\
\hline
\hline
& & & & & &\\[-1em]
41 & $1 / (2 \pi)$ & -0.0605 & 1.8(1.8)  & & 0.36 & 1 \\
31 & "             & "       & 1.9(8)    & & 0.21 & 2 \\
26 & "             & "       & 1.9(6)    & & 0.14 & 3 \\
21 & "             & "       & 1.9(3)    & & 0.11 & 4 \\
15 & "             & "       & 1.9(3)    & & 0.10 & 5 \\
13 & "             & "       & 1.71(15)  & & 0.20 & 6 \\
11 & "             & "       & 1.70(11)  & & 0.17 & 7 \\
10 & "             & "       & 1.9(7)    & & 1.02 & 8 \\
9  & "             & "       & 2.0(6)    & & 1.23 & 9 \\
\hline
\hline
& & & & & &\\[-1em]
11 & " & " & 2.0(6) & -4(7) & 0.14 & 6 \\
10 & " & " & 1.3(4) & 7(5)  & 0.9  & 7 \\
9  & " & " & 1.0(4) & 10(4) & 0.92 & 8 \\
\hline
\end{tabular}
\end{center}
\caption{Summary of the fit systematics for the determination of the large-$N$ behavior
of $\xi^2 \chi$ using the fit function $N\xi^2\chi = e_1 +
e_2/N + e_3/N^2 + e_4/N^3$. Blank spaces mean that the corresponding coefficient
was set to 0 in the fit procedure, while numerical values with no error mean
that the corresponding coefficient was fixed to that value.}
\label{table_large_N_chi_vs_1_on_N}
\end{table}

\subsection{Large-$N$ scaling for $b_2$ and $b_4$}

\begin{figure}[!htb]
\centering
\includegraphics[scale=0.4]{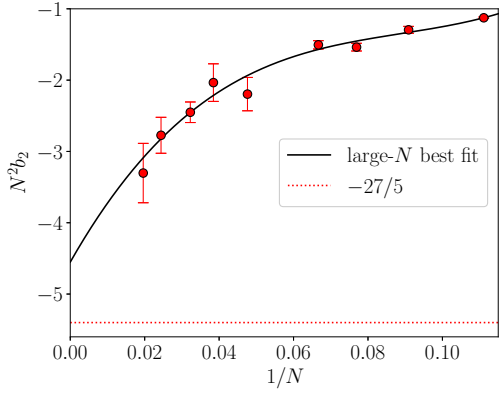}
\caption{Behavior of $N^2 b_2$ as a function of $1/N$ compared to the LO (dashed line) analytic computation of the $1/N$ series. The solid line is the result of a best fit to data with $N \geq 9$ where all coefficients up to the N$^3$LO are fitted (see Tab.~\ref{table_b_2_large_N} for the complete systematics).}
\label{fig_grande_N_b2}
\end{figure}

\begin{figure}[!htb]
\centering
\includegraphics[scale=0.4]{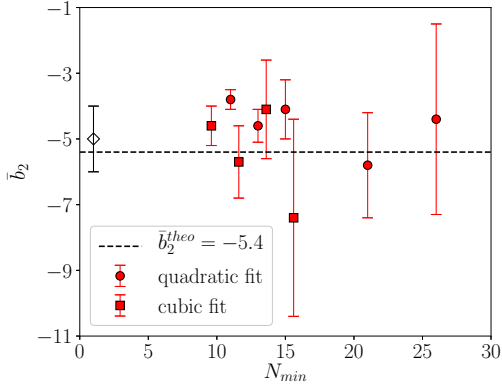}
\caption{Summary of the fit systematics for the LO coefficient $\bar{b}_2$ of the $1/N$ expansion of $b_2$. The empty diamond point represents our overall estimation: $\bar{b}_2=-5(1)$.}
\label{fig_bar_b2}
\end{figure}

We turn now to the analysis of  the large-$N$ limit for $b_2$. In this case, following again the lines of Ref.~\cite{Bonanno:2018xtd}, we employ a fit function including N$^3$LO corrections
\beq\label{fit_function_large_N_b2}
N^2 b_2 = \bar{b}_2 + \frac{k_1}{N} + \frac{k_2}{N^2} + \frac{k_3}{N^3} \, ;
\eeq
that was the minimal polynomial in $1/N$ capable to fit the data available
in Ref.~\cite{Bonanno:2018xtd} after
fixing the leading order term to the predicted theoretical value,
$\bar{b}_2 = - 27/5$. 
In Tab.~\ref{table_b_2_large_N} we report systematics for the best-fit 
procedure, while in Fig.~\ref{fig_grande_N_b2} we plot the best fit 
obtained fitting data to
Eq.~(\ref{fit_function_large_N_b2}) in the whole available range; the systematics for $\bar{b}_2$ are also reported, to improve
clarity, in 
Fig.~\ref{fig_bar_b2}.  
As it can be appreciated, the new data collected at large $N$ permit
us to perform a best fit without fixing the value 
of the LO term. Results obtained for $\bar b_2$
are not stable as the fit range is changed if only NLO corrections
are included in the fit ($k_2 = k_3 = 0$), and tend to be more
and more compatible with the analytic prediction $\bar{b}_2 = -5.4$
as the range is restricted to larger and larger values of $N$.
When $k_2$ and/or $k_3$ are included, results for $\bar{b}_2$ 
are more stable and always compatible with the analytic
prediction: an independent conservative estimate in this case 
would be $\bar{b}_2=-5(1)$.

Present precision allows also to obtain a rough estimation of the order 
of magnitude of the corrections to the large-$N$ scaling of $b_2$: a conservative estimate for the NLO term, which is sufficiently
stable in all performed fits, is
$k_1 = 120(60)$. Our results point out that they seem to increase 
by around one order of magnitude at each step in the expansion, 
resulting in a very slow convergence towards the large-$N$ limit, 
as it has already been observed for the topological susceptibility.
\begin{table}[!t]
\begin{center}
\begin{tabular}{ | c | c | c | c | c | c | c | } 
\hline
& & & & & &\\[-1em]
$N_{\text{\textit{min}}}$ & $\bar{b}_2$ & $k_1$ & $k_2$ & $k_3$ & $\tilde \chi^2$ & dof \\
\hline
\hline
& & & & & & \\[-1em]
41 & -5.5(2.4) & 111(101) & & &  -   & 0 \\
31 & -4.2(8)   & 55(27)   & & & 0.32 & 1 \\
26 & -4.3(7)   & 58(21)   & & & 0.17 & 2 \\
21 & -3.5(4)   & 32(12)   & & & 0.90 & 3 \\
15 & -3.44(21) & 29(3)    & & & 0.70 & 4 \\
13 & -3.09(18) & 21(3)    & & & 2.88 & 5 \\
\hline
\hline
& & & & & & \\[-1em]
26 & -4.4(2.9) & 67(200)  & -156(3000)  & & 0.35 & 1 \\
21 & -5.8(1.6) & 167(100) & -1901(1300) & & 0.33 & 2 \\
15 & -4.1(9)   & 60(40)   & -320(400)   & & 0.75 & 3 \\
13 & -4.6(5)   & 87(19)   & -606(180)   & & 0.69 & 4 \\
11 & -3.8(3)   & 50(10)   & -245(80)    & & 1.56 & 5 \\
9  & -3.56(25) & 40(6)    & -161(40)    & & 1.56 & 6 \\
\hline
\hline
& & & & & & \\[-1em]
15 & -7.4(3.0) & 329(230) & -7020(6000)  & 51526(40000)   & 0.43 & 2 \\
13 & -4.1(1.5) & 50(100)  & 123(1000)    & -4434(11000)   & 0.89 & 3 \\
11 & -5.7(1.1) & 159(60)  & -2133(1000)  & 10022(6000)    & 1.21 & 4 \\
9  & -4.6(6)   & 92(30)   & -939(400)    & 3521(1900)     & 1.25 & 5 \\
\hline
\hline
& & & & & & \\[-1em]
41 & $-27/5$ & 108(9)  & &            & 0.001 & 1 \\
31 & "       & 94(4)   & &            & 1.22  & 2 \\
26 & "       & 93(3)   & &            & 1.07  & 3 \\
21 & "       & 84(3)   & &            & 5.26  & 4 \\
\hline
\hline
& & & & & & \\[-1em]
31 & "       & 147(40) & -1727(1100) & & 0.15 & 1 \\
26 & "       & 134(25) & -1256(800)  & & 0.23 & 2 \\
21 & "       & 144(14) & -1598(400)  & & 0.24 & 3 \\
15 & "       & 121(6)  & -942(100)   & & 1.11 & 4 \\
13 & "       & 117(5)  & -871(60)    & & 1.06 & 5 \\
11 & "       & 98(3)   & -591(40)    & & 5.22 & 6 \\
\hline
\hline
& & & & & & \\[-1em]
26 & "       & 180(110) & -4000(7000) & -50000(100000) & 0.27 & 1 \\
21 & "       & 126(60)  & -573(4000)  & -13634(40000)  & 0.32 & 2 \\
15 & "       & 173(30)  & -3254(1300) & 23117(13000)   & 0.44 & 3 \\
13 & "       & 134(12)  & -1467(400)  & 4959(3500)     & 0.85 & 4 \\
11 & "       & 145(9)   & -1892(260)  & 8768(1700)     & 0.98 & 5 \\
9  & "       & 132(7)   & -1513(150)  & 6040(800)      & 1.36 & 6 \\
\hline
\end{tabular}
\end{center}
\caption{Summary of the fit systematics for the determination of the large-$N$ behavior of $b_2$ using the fit function $N^2 b_2 = \bar{b}_2 +
k_1/N + k_2/N^2+k_3/N^3$. The convention is the same of Tab.~\ref{table_large_N_chi_vs_1_on_N}.}
\label{table_b_2_large_N}
\end{table}
\begin{figure}[!b]
	\centering
	\includegraphics[scale=0.45]{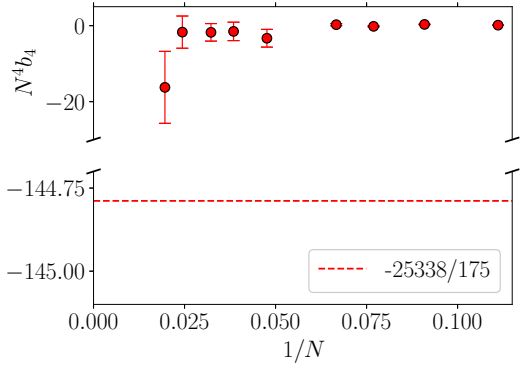}
	\caption{Behavior of $N^4 b_4$ as a function of $1/N$ compared to the LO (dashed line) analytic computation of the $1/N$ series.}
	\label{fig_grande_N_b4}
\end{figure}

Concerning $b_4$, so far continuum extrapolated results,
which are reported in Fig.~\ref{fig_grande_N_b4} in terms 
of the quantity $N^4 b_4$,
are compatible 
with zero within statistical and systematic uncertainties, 
so that we can only set upper bounds.
Nevertheless, such upper bounds
are already interesting enough when compared to LO large-$N$ 
prediction. Indeed, with present data 
one would naively set an upper bound on the modulus of the  
LO order coefficient $ |\bar{b}_4| \lesssim 20$, which is almost 
one order of magnitude smaller than the theoretical prediction,
$|\bar{b}_4|  = |-25338/175| \simeq 145$.
Therefore, we conclude that large corrections 
to LO large-$N$ scaling are expected also in the 
case of $b_4$.

\section{Conclusions}
\label{section_conclusions}

The main purpose of our study was to clarify the matching between
lattice computations and analytic large-$N$ predictions regarding
the dependence on the $\theta$-parameter of 
2$d$ $CP^{N-1}$ models. The picture emerging from previous 
lattice studies pointed out to an apparent disagreement for the sign 
of the deviation of the topological susceptibility
from its LO $1/N$ prediction,
and to values for the LO $1/N^2$ behavior of the $b_2$ coefficient
missing the predicted analytic value by around a factor 2. 
A possible way out was proposed 
in Ref.~\cite{Bonanno:2018xtd}, which showed that assuming 
large higher order contributions in the $1/N$ expansion, 
the disagreement could disappear, leading at least to a 
partial consistency between lattice data and analytic 
computations. In Ref.~\cite{Bonanno:2018xtd} it was also pointed
out that, assuming the presence of such higher-order corrections,
it was necessary to reach at least $N = 50$ to make the situation
clearer.

In this work we accomplished this goal, exploiting 
a recent algorithm proposed by 
M.~Hasenbusch
to defeat critical slowing down~\cite{fighting_slowing_down_cpn},
which has been adapted to our Symanzik improved discretization in the 
presence of a $\theta$-term. 
That allowed us to push our investigation up to $N = 51$.

In this way we have been able to provide independent determinations of 
the NLO
and LO coefficients, respectively for $\chi$ and $b_2$, 
which are in agreement with analytic predictions.
In particular, we have estimated
$e_2=-0.05(3)$ (analytic prediction $e_2\simeq-0.0605$~\cite{calcolo_e_2})
and $\bar{b}_2=-5(1)$ (analytic prediction $\bar{b}_2 = -5.4$). At the same time, we have provided a first estimate for the NNLO contribution
to $\chi$ ($e_3=1.5(5)$) and for the NLO contribution to $b_2$
($k_1 \sim 120(60)$) which are presently unknown analytically.

Instead, no definite results have been obtained for $b_4$, because of the 
larger statistical uncertainties involved in the determination
of this observable, apart from upper bounds which 
however seem to be in disagreement with the LO large-$N$ prediction
by around one order of magnitude, pointing to the presence 
of large NLO corrections also in this case.

Our results, apart from successfully confirming present analytic
estimates for $\chi$ and $b_2$, point out that the convergence of 
the large-$N$ expansion is particularly slow for this class of models.
As suggested in Ref.~\cite{Bonanno:2018xtd}, this could be due
to the non-analytic behavior which is expected for $N = 2$.

\acknowledgments
The authors thank C.~Bonati, M.~Hasenbusch, P.~Rossi and E.~Vicari for useful discussions. 

Numerical simulations have been performed at the Scientific Computing Center at INFN-PISA and on the MARCONI machine at CINECA, based on the agreement between INFN
and CINECA (under projects INF18\_npqcd and INF19\_npqcd).

\end{document}